\documentclass[aps,prb,reprint,groupedaddress,amsmath,amssymb]{revtex4-1}
\usepackage{bm}
\usepackage{braket}
\usepackage{graphicx,color}

\begin{document}

\title{Interaction effect on reservoir-parameter-driven adiabatic charge pumping via a single-level quantum dot system}
\author{Masahiro Hasegawa}
\author{Takeo Kato}
\affiliation{Institute for Solid State Physics, The University of Tokyo, Kashiwa, Chiba 277-8581, Japan}

\date{\today}

\begin{abstract}
We formulate adiabatic charge pumping via a single-level quantum dot (QD) induced by reservoir parameter driving, i.e., temperature and electrochemical potential driving.
Our formulation describes arbitrary strength of dot-reservoir coupling and Coulomb interaction in the QD, and is applicable to the low-temperature regime, where the Kondo effect becomes important.
The adiabatic charge pumping is expressed by the Berry connection and is related to delayed response of the QD.
We calculate the pumped charge by the renormalized perturbation theory, and discuss how the Coulomb interaction affects the charge pumping.
\end{abstract}


\maketitle

\section{Introduction}
Recent development of nanotechnology has been enabled us to control and measure nanoscale quantum devices with high accuracy.
Time-dependent transport of nanoscale devices under bias parameter driving has attracted much interest, and has been investigated for, e.g, single-electron current source~\cite{Kouwenhoven91,Pekola13a,Connolly13,Roche13} and single electron generator~\cite{Feve07,Bocquillon13,Dubois13}.
Development of novel quantum devices usually aims to improve their function, e.g., accurate control and reliable evaluation of a generated  current.
However, in future, suppression of undesirable energy losses will be important to improve efficiency of quantum devices.
This problem can be discussed in the context of quantum thermodynamics by regarding the parameter driving as a cyclic operation on small systems.
Actually, efficiency and power of cyclic operations have been discussed energetically for, e.g., finite-time thermodynamics~\cite{Chambadal57,Novikov58,Curzon75,Schmiedl08,Anderson11,Seifert11,Whitney14,Shiraishi16}, quantum engines~\cite{Henrich07,Esposito10}, steady-state thermodynamics~\cite{Oono97,Komatsu10,Saito11,Yuge13,Taguchi16}, and information feedback~\cite{Sagawa10,Toyabe10,Parrondo15}.

Adiabatic pumping in nanoscale circuits is one of basic concepts in study of quantum thermodynamics.
Adiabatic charge pumping was first proposed by Thouless~\cite{Thouless83}, and was formulated for mesoscopic systems by B\"uttiker \textit{et. al.}~\cite{Buttiker93,Buttiker94,Pretre96}.
Brouwer's formula~\cite{Brouwer98} is a milestone, which rewrites the previous theoretical formulation of adiabatic pumping into a simple form by the Berry connection.
These theoretical works, however, discussed only noninteracting electron systems or interacting electron systems within a mean-filed theory.
After these works, adiabatic pumping in interacting electron systems has been investigated in several regimes, such as the Coulomb blockage regime~\cite{Aleiner98,Brouwer05}, the weak interaction regime~\cite{Hernandez09}, the incoherent transport regime~\cite{Splettstoesser06,Reckermann10,Calvo12} and the Kondo regime~\cite{Aono04,Splettstoesser05,Sela06,Eissing16,Romero17}.
The Kondo effect on time-dependent transport is one of the most important and challenging problems, and has been discussed for a long time by several theoretical approaches, e.g., non-crossing approximation~\cite{Hettler95}, equation of motion approach~\cite{Ng96}, perturbation theory with respect to coupling~\cite{Kaminski00}, slave-boson mean-field theory~\cite{Aono04,Splettstoesser05}, scattering theory~\cite{Sela06}, renormalization group method~\cite{Eissing16} and numerical renormalization group calculation~\cite{Romero17}.

\begin{figure}[bt]
\begin{center}
\includegraphics[width=5.5cm]{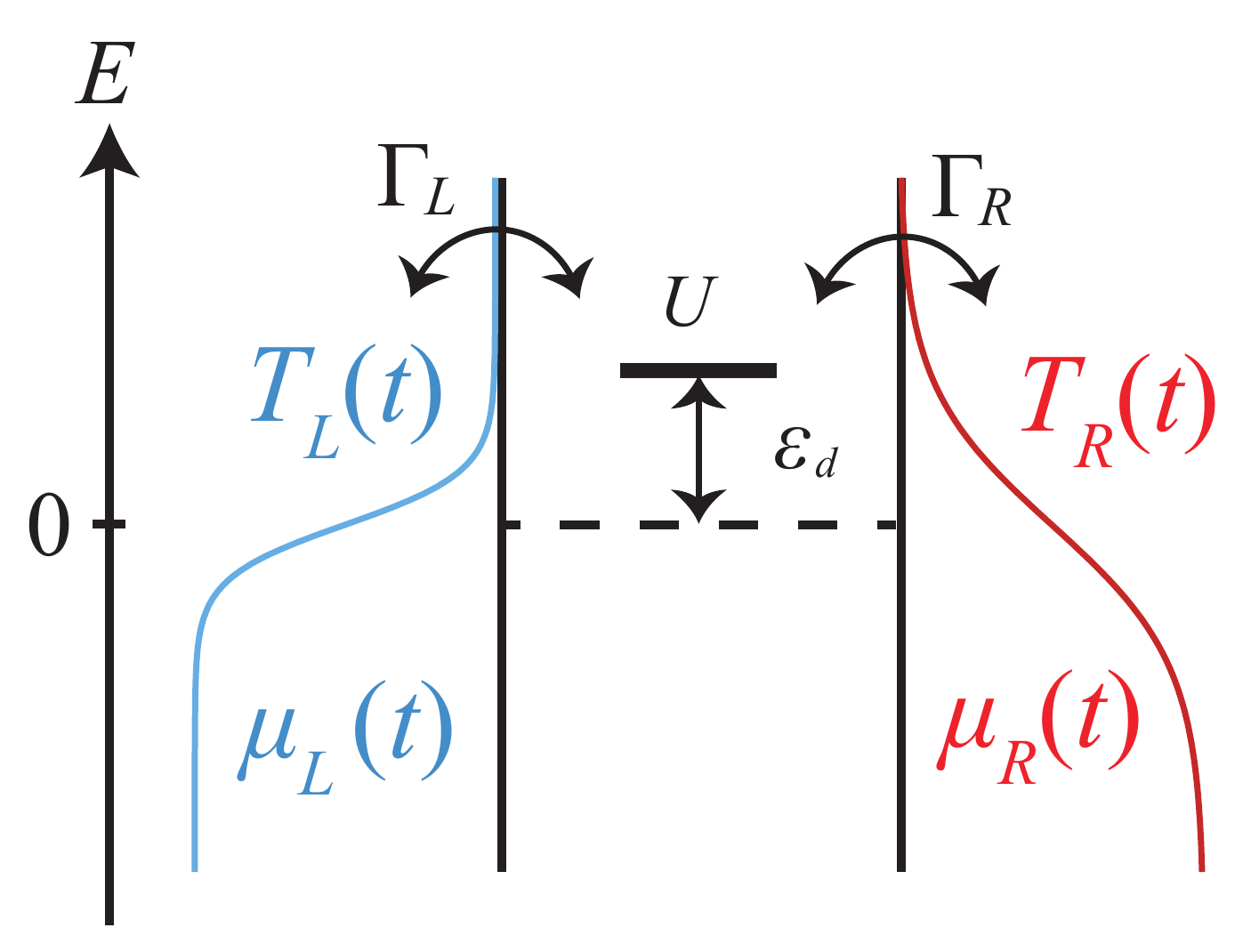}
\caption{\label{fig:dot_pic}
A schematic of our model. A single-level QD with two electron reservoirs.
$\epsilon_d$ and $U$ denotes an energy level of the QD and the Coulomb interaction within the QD, respectively.
Subscripts ${\cdot}_L$ and ${\cdot}_R$ denotes reservoir index and $\Gamma_{L/R}$, $T_{L/R}(t)$ and $\mu_{L/R}(t)$ denotes linewidth and time dependent reservoir temperatures and electrochemical potentials, respectively.
}
\end{center}
\end{figure}

The previous theoretical works have mainly focused on adiabatic pumping induced by gate voltages or voltage biases.
Even though temperature driving is one of the basic operations in thermodynamics to discuss efficiency of quantum devices, it has been studied only in a few theoretical works.
Heat pumping via molecular junction~\cite{Ren10} and excess entropy production in steady-state system~\cite{Yuge13,Taguchi16} are examples of such works. 
These works are based on generalized quantum Master equation approach~\cite{Schoeller94} to describe time-dependent temperature.
However, since the generalized quantum Master equation describes transport up to lowest-order perturbation of system-reservoir coupling, it cannot describe low-temperature regime, where effect of higher-order perturbation becomes significant.
In order to understand thermodynamics for small quantum devices, it is unavoidable to construct a formula for time-dependent transport induced by temperature driving, which is applicable even to the low-temperature region.
In addition, it is also a challenging and important problem to study many-body effects, especially, the Kondo effect which is remarkable in quantum dot (QD) systems with the strong Coulomb interaction at low temperatures.

In our previous paper~\cite{Hasegawa17}, we have formulated the adiabatic charge pumping via a single-level QD induced by time-dependent reservoir parameters, i.e., temperatures and electrochemical potentials for arbitrary QD-reservoir coupling strength within the first-order perturbation with respect to the Coulomb interaction of the QD (Fig.~\ref{fig:dot_pic}).
In this paper, we generalize our previous theoretical work to discuss many-body effect for arbitrary strength of the Coulomb interaction.
We note that our formalism is similar to that of the recent theoretical work~\cite{Romero17} though only electrochemical potential driving has been considered.
Using the Keldysh formalism and the Ward identity, we formulate adiabatic charge pumping for arbitrary strength of the Coulomb interaction in the QD, and express it by the Berry connection.
To describe effect of time-dependent reservoir temperatures, we employ thermomechanial field method~\cite{Eich14,Tatara15}.
To clarify physical meaning of the Berry connection, we discuss delayed response of the QD, and relate it to dynamic AC response.
Finally, we calculate pumped charges by using the renormalized perturbation theory (RPT)~\cite{Hewson93,Hewson01}.

This paper is organized as follows.
In Sec.~\ref{sec:formalism}, we introduce our Hamiltonian, define Keldysh Green's functions, and give a brief introduction of the thermomechanical field method.
In Sec.~\ref{sec:general_formalims}, we derive our general formula for the pumped charge by the adiabatic approximation.
In Sec.~\ref{sec:mechanism}, we discuss mechanism of the present pumping in terms of response delay time.
In Sec.~\ref{sec:application}, we presents the results for both the electrochemical-potential-driven case and the temperature-driven case obtained by RPT.
Finally, we summarize our results In Sec.~\ref{sec:summary}.
In the Appendices, we present details for derivation of certain relevant equations.

\section{Formalism \label{sec:formalism}}
In this section, we formulate pumping current induced by time-dependent reservoir parameters.
After introducing a model of a single-level QD connected with electron reservoirs with time-dependent parameters (Sec.~\ref{sec:Hamiltonian}),
we define several Keldysh Green's functions (Sec.~\ref{sec:KeldyshGF}), and introduce a thermomechanical field, which is an artificial field to describe time-dependent reservoir temperatures (Sec.~\ref{sec:TMFmethod}).
Finally, we describe the charge current in terms of Keldysh Green's functions.
Throughout this paper, we employ the unit system of $\hbar = k_B = 1$, and define the electron charge as $e < 0$.

\subsection{Model\label{sec:Hamiltonian}}

We consider the Anderson impurity model~\cite{Anderson61} with time-dependent reservoir temperatures and electrochemical potentials, whose Hamiltonian is given by
\begin{align}
	H &= H_d + \sum_{r=L,R}H_r +  H_T, 
\end{align}
where $H_d$, $H_r$, and $H_T$ describe the QD, the electron reservoir $r \in \{ L, R \}$ and a QD-reservoir coupling, respectively:
\begin{align}
	H_d &= \epsilon_d d_s^{\dagger} d_s + U d_{\uparrow}^{\dagger} d_{\uparrow} d_{\downarrow}^{\dagger} d_{\downarrow} , \\
	H_r &= \sum_{k,s} ( B_r(t) \epsilon_{k} + \mu_r(t) )c_{rks}^{\dagger} c_{rks} , \label{eqn:reservoir_hamiltonian}\\
	H_T &= v_r(t) c_{rks}^{\dagger} d_s + h.c. 
\end{align}
Here, $d_s^{\dagger}\ (d_s)$ denotes a creation (annihilation) operator of an electron in the QD with a spin $s \in \{ \uparrow, \downarrow \}$, and $c_{rks}^{\dagger}\ (c_{rks})$ denotes that of an electron in the reservoir $r$ with a spin $s$ and a wavenumber $k$.
The electron energies in the QD and the reservoirs are denoted by $\epsilon_d$ and $\epsilon_{k}$, respectively, and $U$ is a strength of the Coulomb interaction in the QD.
We have introduced an electrochemical potential $\mu_r(t)$ and a thermomechanical field $B_r(t)$ to describe the time-dependent reservoir parameters. 
The thermomechanical field describes time-dependent temperature and its theoretical mechanism is explained in Sec.~\ref{sec:TMFmethod}.
We assume that they are periodic functions of $t$ as
\begin{align}
	B_r(t) =  B_r(t + \mathcal{T}) , \quad \mu_r(t) =  \mu_r(t + \mathcal{T}),
\end{align}
where $\mathcal{T}$ is a period of pumping.
In the presence of the thermomechanical field, the QD-reservoir coupling in $H_T$ have to be taken as
\begin{align}
	v_r(t) = v_r \sqrt{B_r(t)} , \label{eqn:time_dependent_coupling}
\end{align}
where $v_r$ is a time-independent coupling constant (for details, see Sec.~\ref{sec:TMFmethod}).
For simplicity, the chemical potentials of the reservoirs is set to zero in the absence of the parameter driving ($B_r(t) = 1$, $\mu_r(t) = 0$).
We also assume that, without the parameter driving, the reservoirs are in thermal equilibrium with the reference temperature $T$.

Throughout this paper, we consider the wide-band limit, in which the sum with respect to $k$ is converted into the energy integral as
\begin{align}
	\sum_k ( \cdots ) \simeq \rho \int_{-\infty}^{\infty} d\epsilon_k ( \cdots ) \ ,
\end{align}
where $\rho$ is the density of states at the Fermi level of the reservoirs.
In this limit, a linewidth of the QD is defined as
\begin{align}
	\Gamma &= \sum_{r=L,R} \Gamma_r , \\
	\Gamma_r &= 2\pi \rho |v_r|^2 .
\end{align}

\subsection{Keldysh Green's functions\label{sec:KeldyshGF}}

We define GFs and self-energies as a function of time variables on the Keldysh contour, $C_K$, and project them as necessary onto two real-time contours, the forward contour $C_+$ and the backward contour $C_-$~\cite{Keldysh64}.
Hereafter, time variables on the Keldysh contour are denoted by the Greek characters (e.g., $\tau$), and those on the real-time contour by the italic characters (e.g., $t$).

We define a (full) GF of an electron in the QD as
\begin{align}
	G_{s}(\tau_1,\tau_2) = -i \Braket{T_K d_s(\tau_1) d_s^{\dagger}(\tau_2)},
\end{align}
where $T_K$ is a time ordering operator on the Keldysh contour.
The Dyson's equation for this GF is written as
\begin{align}
	G_s(\tau_1,\tau_2) &= g_s(\tau_1,\tau_2)  \nonumber \\
	& \hspace{-8mm} + \int_{C_K} d\tau_3 d\tau_4 \ g_s(\tau_1,\tau_3) \Sigma_{s}(\tau_3,\tau_4) G_s(\tau_4,\tau_2),
\end{align}
where $g_s(\tau_1,\tau_2)$ and $g_{rks}(\tau_1,\tau_2)$ is noninteracting GFs of an electron for the isolated QD and the isolated reservoirs defined by
\begin{align}
	g_s(\tau_1,\tau_2) &= -i \Braket{T_K d_s(\tau_1) d_s^{\dagger}(\tau_2)}_{v_r=U=0} , \\
	g_{rks}(\tau_1,\tau_2) &= -i \Braket{T_K c_{rks}(\tau_1) c_{rks}^{\dagger}(\tau_2)}_{v_r=U=0} ,
\end{align}
respectively, and $\Sigma_s(\tau_3,\tau_4)$ is a one-particle-irreducible (1PI) self-energy.
The 1PI self-energy is composed of two terms as
\begin{align}
	\Sigma_{s}(\tau_1,\tau_2) = \Sigma_{0,s}(\tau_1,\tau_2) + \Sigma_{U,s}(\tau_1,\tau_2) ,
\end{align}
where $\Sigma_{0,s}(\tau_1,\tau_2)$ and  $\Sigma_{U,s}(\tau_1,\tau_2)$ denote self-energies due to the QD-reservoir coupling and the Coulomb interaction in the QD, respectively.
The former is described simply by the isolated-reservoir GF, $g_{rks}$, as
\begin{align}
	\Sigma_{0,s}(\tau_1,\tau_2) &= \sum_{r=L,R} \Sigma_{0,s,r}(\tau_1,\tau_2) , \\
	\Sigma_{0,s,r}(\tau_1,\tau_2) &= \sum_k v_r^{\ast}(\tau_1) g_{rks}(\tau_1,\tau_2) v_r(\tau_2) .
\end{align}
Hereafter, we call $\Sigma_{0,s}(\tau_1,\tau_2)$ as the reservoir self-energy and $\Sigma_{U,s}(\tau_1,\tau_2) $ as the interaction self-energy.

\subsection{Thermomechanical field method\label{sec:TMFmethod}}

The thermomechanical field method was first proposed by Luttinger~\cite{Luttinger64}, and has been employed in recent theoretical works~\cite{Eich14,Tatara15,Hasegawa17}.
In this section, we briefly explain how the thermomechanical field modifies the reservoir temperature.
For detailed discussion, see Ref.~\onlinecite{Hasegawa17}.

\begin{figure}[tb]
\begin{center}
\includegraphics[width=6.5cm]{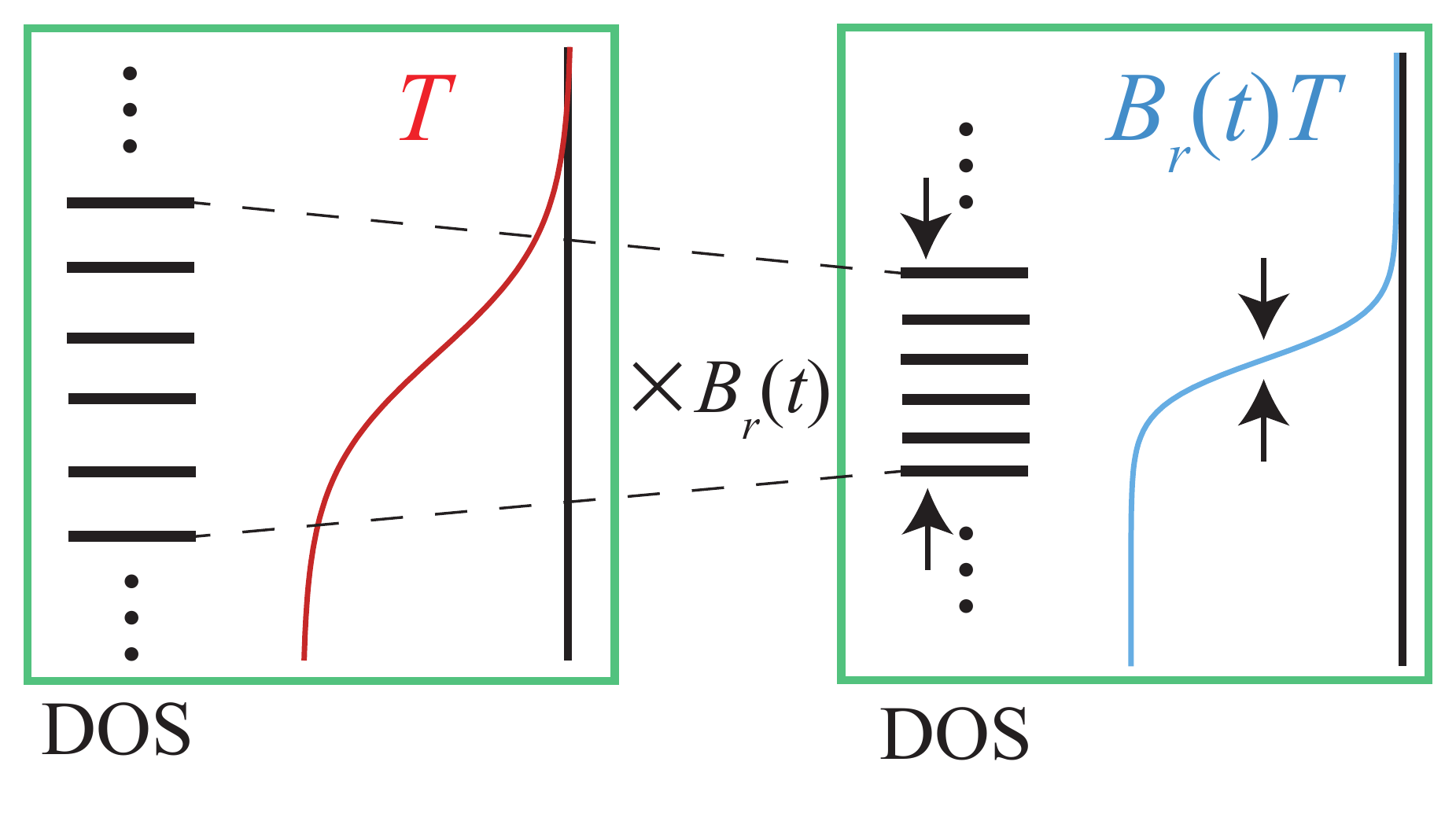}
\caption{\label{fig:tmf}
A schematic of energy rescaling induced by the thermomachanical field.
The left panel shows reservoir energy levels and the Fermi distribution for the original setup ($B_r(t) = 1$), and the right panel shows those modified by a thermomechanical field for $B_r(t) <1$.
By this energy rescaling, the reservoir temperature effectively decreases.
}
\end{center}
\end{figure}

Let us consider the case of $\mu_r(t)=0$ for simplicity.
As seen in the Hamiltonian given by Eq.~(\ref{eqn:reservoir_hamiltonian}), the thermomechanical field rescales electron energies in the reservoirs as $\epsilon_k \rightarrow B(t) \epsilon_k$.
Fig.~\ref{fig:tmf} is a schematic of how this energy rescaling modifies the reservoir temperature.
First, before the energy scaling ($B_r(t)=1$), the electron reservoirs are both in thermal equilibrium with reference temperature $T$ and their Fermi distribution function is denoted by $f_r(\epsilon) = f(\epsilon,T,0)$, where $f(\epsilon,T,\mu) = [ e^{(\epsilon -\mu )/  T} + 1]^{-1}$.
Then, after the energy scaling ($B_r(t)\neq1$), each Fermi distribution function of the reservoirs are rescaled into $\tilde{f}_r(\epsilon)$ as
\begin{align}
	\tilde{f}_r(B_r(t)\epsilon) = f_r(\epsilon) ,
\end{align}
and this energy rescaling can be put onto temperature directly as
\begin{align}
	\tilde{f}_r(\epsilon) = f_r(\epsilon / B_r(t)) = f(\epsilon,B_r(t)T,0) .
\end{align}
As a result, the reservoir temperature is rescaled by the thermomechanical field as
\begin{align}
	T_r(t) = B_r(t) T . \label{eqn:time_dependent_temperature}
\end{align}
To keep the temperature positive, the thermomechanical field $B(t)$ have to be chosen as positive.

We make a remark on this thermomechanical field method.
The thermomechanical field rescales not only the temperature but also the linewidth $\Gamma$.
To cancel this undesirable effect, we have to adjust the QD-reservoir coupling in an appropriate way.
Actually, if we introduce the time-dependent QD-reservoir coupling given in Eq.~(\ref{eqn:time_dependent_coupling}), the linewidth is shown to remain constant for arbitrary forms of $B(t)$ (see also Appendix~\ref{app:adiabtic_selfenergy}).
We also note that the present method can be applied only to continuous and infinite reservoir systems.
This condition is satisfied if we take the wide-band limit for the reservoirs.

\subsection{Charge current}

A charge current flowing from the reservoir $r$ into the QD at time $t$ is defined as 
\begin{align}
	J_r(t) =  - e \frac{d}{dt}\Braket{\sum_{k,s} c_{rks}^{\dagger}(t) c_{rks}(t)}.
\end{align}
By the Heisenberg equation of motion, the charge current is rewritten as
\begin{align}
	J_r(t) = 2e \sum_{k,s} \mathrm{Re} \left[ v_r(t) G_{s;rks}^{<}(t,t) \right] , \label{eqn:current_before_ket}
\end{align}
where $G_{s;rks}^<(t_1,t_2)$ is a lesser component of the Keldysh GF defined by
\begin{align}
	G_{s;rks}^<(t_1,t_2) = i \Braket{c_{rks}^{\dagger}(t_2) d_s(t_1)} .
\end{align}
By the technique of the kinetic equation and Langreth's theorem~\cite{Jauho94}, the current is expressed
only by the GFs and the self-energies of the QD as
\begin{align}
	J_r(t) &= 2e \sum_s \int dt_1 \mathrm{Re} \Bigl[  G_s^{R}(t,t_1) \Sigma_{0,s,r}^<(t_1,t)  \nonumber \\
	& \hspace{28mm} + G_s^{<}(t,t_1) \Sigma_{0,s,r}^A(t_1,t) \Bigr] . \label{eqn:charge_current}
\end{align} 
Here, $R$ and $A$ in the superscript indicate a retarded and an advanced component, respectively.

\section{Adiabatic Approximation\label{sec:general_formalims}}

In this section, we derive expressions of the pumping current in the adiabatic approximation.
Throughout this section, we use a parameter vector ${\bm X}(t)$ defined by
\begin{align}
	\bm{X}(t) = \Bigl(X_\mu(t)\Bigr) = \Bigl(T_L(t),T_R(t),\mu_L(t),\mu_R(t)\Bigr)^T,
    \label{eq:defXvector}
\end{align}
to simplify the equations.

\subsection{\label{sec:adiabatic_approximation}Outline}

Let us consider the adiabatic process, in which the reservoir parameters are driven sufficiently slow so that higher-order time derivatives of time-dependent parameters can be neglected.
Although a standard procedure for the adiabatic approximation utilizes the gradient expansion~\cite{Rammer86}, we employ a more heuristic method here.
We first expand the reservoir parameters around the time $t$ as
\begin{align}
	B_r(t_1) &\simeq B_r(t) + \dot{B}_r(t)(t_1-t) ,  \label{eqn:add_app_tmf} \\
	\mu_r(t_1) &\simeq \mu_r(t) + \dot{\mu}_r(t)(t_1-t) . \label{eqn:add_app_mu}
\end{align}
For simplicity of description, we rewrite Eqs.~(\ref{eqn:add_app_tmf}) and (\ref{eqn:add_app_mu}) with the parameter vector $\bm{X}(t)$ as
\begin{align}
	X_\mu(t_1) &\simeq X_\mu(t) + \dot{X}_\mu(t)(t_1-t), \hspace{3mm} (\mu=1,2,3,4).
\end{align}
The reference time $t$ is taken as the time of the observables to be measured since the observables should depend on $X_\mu(t)$ and $\dot{X}_\mu(t)$ in the adiabatic approximation.
In this approximation, the charge current $J_r(t)$ is given by a sum of two terms as
\begin{align}
	& J_r(t) \simeq J_r^{\mathrm{st.}}({\bm X}(t)) +  J^{\mathrm{ad.}}_r({\bm X}(t),\dot{\bm X}(t)),
\end{align}
where
\begin{align}
	& J^{\mathrm{st.}}_r({\bm X(t)}) =  \left. J_r(t) \right|_{\dot{\bm X}(t)={\bm 0}} , \label{eqn:def_of_steady_term} \\
	& J^{\mathrm{ad.}}_r({\bm X}(t),\dot{\bm X}(t)) = \sum_{\mu=1}^{4} \left. \frac{\partial J_r(t)}{\partial \dot{X}_{\mu}} 
	\right|_{\dot{\bm X}(t)={\bm 0}} \dot{X}_{\mu}(t) .
\end{align}
Here, $J^{\mathrm{st.}}_r({\bm X}(t))$ is a steady-state current for the fixed reservoir parameters, and is independent of the frequency of the parameter driving.
On the other hand, $J^{\mathrm{ad.}}_r({\bm X}(t),\dot{\bm X}(t))$ is an adiabatic correction, which generates the charge current proportional to the pumping frequency.

The purpose of this section is to derive formulas for $J_r^{\mathrm{st.}}({\bm X}(t))$ and $J^{\mathrm{ad.}}_r({\bm X}(t),\dot{\bm X}(t))$ from the definition of the charge current, Eq.~(\ref{eqn:charge_current}).
For this purpose, we first consider the adiabatic approximation for the self-energies and the GFs in the subsequent two subsections.

\subsection{Self-energies}

The retarded and advanced components of the self-energies are calculated for arbitrary driving of reservoir parameters as
\begin{align}
	\Sigma_{0,s,r}^R(t_1,t_2) &= -i \frac{\Gamma_r}{2} \delta(t_1-t_2) , \label{eqn:retarded_reservoir_selfenergy} \\
	\Sigma_{0,s,r}^A(t_1,t_2) &= i \frac{\Gamma_r}{2} \delta(t_1-t_2) . \label{eqn:advanced_reservoir_selfenergy}
\end{align}
Details of derivation is given in Appendix~\ref{app:adiabtic_selfenergy}.
These components are independent of the reservoir parameters, and remain constant in time evolution.
Therefore, there is no adiabatic correction for these components of the reservoir self-energies.

The lesser component indeed has an adiabatic correction term and is approximated as
\begin{align}
\Sigma_{0,s,r}^<(t_1,t_2) &\simeq  \Sigma_{0,s,r}^{<,\mathrm{st.}}(t_1,t_2;{\bm X}(t)) \nonumber \\
&\hspace{0.5cm} + \Sigma_{0,s,r}^{<,\mathrm{ad.}}(t_1,t_2;{\bm X}(t),\dot{\bm X}(t)) ,
\label{eqn:selfenergy_lesser_adiabatic}
\end{align}
where
\begin{align}
& \Sigma_{0,s,r}^{<,\mathrm{st.}}(t_1,t_2;{\bm X}(t)) = i \Gamma_r \! \int \! \frac{d\epsilon}{2\pi}  f_r(\epsilon, {\bm X}(t)) e^{-i \epsilon (t_1-t_2)} , \label{eqn:selfenergy_lesser_adiabatic} \\
&\Sigma_{0,s,r}^{<,\mathrm{ad.}}(t_1,t_2;{\bm X}(t),\dot{\bm X}(t))= i \Gamma_r \! \int \! \frac{d\epsilon}{2\pi}  f_r(\epsilon, {\bm X}(t)) e^{-i \epsilon (t_1-t_2)} 
\nonumber \\
& \hspace{5mm}\times \biggl\{ \frac{\dot{T}_r(t)}{T_r(t)} [1-i(\epsilon - \mu_r(t))(t_1-t_2)] - i \dot{\mu}_r(t) (t_1-t_2)  \biggr\} \nonumber \\
& \hspace{5mm}\times \biggl[ \frac{t_1+t_2}{2} - t \biggr] ,
\label{eqn:adiabatic_reseroir_lesser_selfenergy}
\end{align}
and $f_r(\epsilon,{\bm X}(t)) = f(\epsilon,T_r,\mu_r)$.
Detailed derivation is given in Appendix~\ref{app:adiabtic_selfenergy}.
We note that Eq.~(\ref{eqn:adiabatic_reseroir_lesser_selfenergy}) can be rewritten as
\begin{align}
& \Sigma_{0,s,r}^{<,\mathrm{ad.}}(t_1,t_2;{\bm X}(t),\dot{\bm X}(t))\nonumber \\
& \hspace{5mm} =\biggl( \frac{t_1+t_2}{2} - t \biggr) \frac{\partial \Sigma_{0,s,r}^{<,\mathrm{st.}}}
{\partial t}(t_1,t_2;{\bm X}(t)),
\end{align}
which is the equation of the adiabatic approximation employed in Ref.~\onlinecite{Splettstoesser05}.
By using this relation, one can check that the present approximation scheme reproduces the results of the adiabatic approximation based on the gradient expansion.

\subsection{\label{sec:Adiabatic_Green's_functions}Green's functions}

Both the retarded and the lesser component of the GFs have the adiabatic correction as
\begin{align}
	G_s^{R}(t_1,t_2) &\simeq G_s^{R,\mathrm{st.}}(t_1,t_2;{\bm X}(t)) \nonumber \\
    &\hspace{0.5cm} + G_s^{R,\mathrm{ad.}}(t_1,t_2;{\bm X}(t),\dot{\bm X}(t)) , \\
	G_s^{<}(t_1,t_2) &\simeq G_s^{<,\mathrm{st.}}(t_1,t_2;{\bm X}(t)) \nonumber \\
    &\hspace{0.5cm} + G_s^{<,\mathrm{ad.}}(t_1,t_2;{\bm X}(t),\dot{\bm X}(t)) . 
\end{align}
The steady-state GFs are easily calculated from the Dyson equation as
\begin{align}
&G_s^{R,\mathrm{st.}}(t_1,t_2;{\bm X}(t)) = \left( G_s^{A,\mathrm{st.}}(t_2,t_1;{\bm X}(t)) \right)^{\ast}  \nonumber \\
& \hspace{3mm}= g_s^{R}(t_1,t_2)  + \int \! dt_3 dt_4 \, g_s^R(t_1,t_3) \Sigma_s^{R,{\rm st.}}(t_3,t_4; {\bm X}(t)) \nonumber \\
& \hspace{50mm} \times  G_s^{R,\mathrm{st.}}(t_4,t_2;{\bm X}(t)), 
\end{align}
and
\begin{align}
&G_s^{<,\mathrm{st.}}(t_1,t_2;{\bm X}(t))  \nonumber \\
& \hspace{5mm} =  \int \! dt_3 dt_4 \, G_s^R(t_1,t_3;{\bm X}(t))\Sigma_s^{<,{\rm st.}}(t_3,t_4; {\bm X}(t)) \nonumber \\
& \hspace{50mm} \times G_s^{A,\mathrm{st.}}(t_4,t_2;{\bm X}(t)),
\end{align}
where 
\begin{align}
& \Sigma_s^{R,{\rm st.}}(t_1,t_2; {\bm X}(t)) \nonumber \\ 
&\hspace{5mm} = \sum_r \Sigma_{0,s,r}^{R,{\rm st.}}(t_1,t_2) + \Sigma_{U,s}^{R,{\rm st.}}(t_1,t_2;{\bm X}(t)), \\
& \Sigma_s^{<,{\rm st.}}(t_1,t_2; {\bm X}(t)) \nonumber \\
&\hspace{5mm} = \sum_r \Sigma_{0,s,r}^{<,{\rm st.}}(t_1,t_2; {\bm X}(t)) + \Sigma_{U,s}^{<,{\rm st.}}(t_1,t_2;{\bm X}(t)), 
\end{align}
We note that the interaction self-energies, $\Sigma_{U,s}^{R,{\rm st.}}$ and $\Sigma_{U,s}^{<,{\rm st.}}$, include the steady-state reservoir self-energies, $\Sigma_{0,s,r}^{R,{\rm st.}}$ and $\Sigma_{0,s,r}^{<,{\rm st.}}$ via internal propagators in their Feynman diagrams.

The adiabatic correction term is the next leading term in the adiabatic approximation.
Thus, all we have to do to obtain the adiabatic correction term of the GF is replacing one of the steady state terms of the reservoir self-energies, $\Sigma_{0,s,r}^{<,\mathrm{st.}}$, into the adiabatic correction term, $\Sigma_{0,s,r}^{<,\mathrm{ad.}}$, in each diagram.
This operation is denoted by functional derivative as
\begin{align}
	&G_s^{R,\mathrm{ad.}}(t_1,t_2;{\bm X}(t),\dot{\bm X}(t))  \nonumber \\
	& \hspace{5mm}= \sum_{s^{\prime}} \int dt_3 dt_4 \ \frac{\delta G_s^{R,\mathrm{st.}}(t_1,t_2;{\bm X}(t))}{\delta \Sigma_{0,s^{\prime}}^{<,\mathrm{st.}}(t_4,t_3;{\bm X}(t))} \nonumber \\
    & \hspace{20mm} \times \Sigma_{0,s^{\prime}}^{<,\mathrm{ad.}}(t_4,t_3;{\bm X}(t),\dot{\bm X}(t)) ,  \label{eqn:retarded_functional_derivative_first} \\
	&G_s^{<,\mathrm{ad.}}(t_1,t_2;{\bm X}(t),\dot{\bm X}(t)) \nonumber \\
	& \hspace{5mm}= \sum_{s^{\prime},r} \int dt_3 dt_4 \ \frac{\delta G_s^{<,\mathrm{st.}}(t_1,t_2;{\bm X}(t))}{\delta \Sigma_{0,s^{\prime}}^{<,\mathrm{st.}}(t_4,t_3;{\bm X}(t))} \nonumber \\ 
    &  \hspace{20mm} \times\Sigma_{0,s^{\prime}}^{<,\mathrm{ad.}}(t_4,t_3;{\bm X}(t),\dot{\bm X}(t)) . \label{eqn:lesser_functional_derivative_first}
\end{align}
These functional derivatives can be rewritten into the GFs and the connected two-particle GFs as
\begin{align}
	\frac{\delta G_s^{R,\mathrm{st.}}(t_1,t_2;{\bm X}(t))}{\delta \Sigma_{0,s^{\prime}}^{<,\mathrm{st.}}(t_4,t_3;{\bm X}(t))} 
    &= \gamma_{s;s^{\prime}}^{R;<,\mathrm{st.}}(t_1,t_2;t_3,t_4;{\bm X}(t)) , \label{eqn:funcder_GF_to_4point_ret} \\
	\frac{\delta G_s^{<,\mathrm{st.}}(t_1,t_2;{\bm X}(t))}{\delta \Sigma_{0,s^{\prime}}^{<,\mathrm{st.}}(t_4,t_3;{\bm X}(t))} &= \gamma_{s;s^{\prime}}^{<;<,\mathrm{st.}}(t_1,t_2;t_3,t_4;{\bm X}(t)) . \label{eqn:funcder_GF_to_4point_les}
\end{align}
Here $ \gamma_{s;s^{\prime}}^{R;<,\mathrm{st.}}$ and $\gamma_{s;s^{\prime}}^{<;<,\mathrm{st.}}$ are functions composed of the GFs and the connected two-particle GFs and its detail formulas are given in Appendix~\ref{app:derivation_ward_identity}.

\subsection{Charge current}

Applying the adiabatic approximation discussed above to the charge current, Eq.~(\ref{eqn:charge_current}), we obtain
\begin{align}
	J_r(t) &\simeq J_r^{\mathrm{st.}}({\bm X}(t)) + J_r^{\mathrm{ad.}}({\bm X}(t),\dot{\bm X}(t)) , \label{eqn:charge_current_two_app}
\end{align}
where
\begin{align}
	& J_r^{\mathrm{st.}}({\bm X}(t))= 2e \sum_s \int \! dt_1 \nonumber \\
    & \hspace{3mm} \times \mathrm{Re} \Bigl[  G_s^{R,\mathrm{st.}}(t,t_1;{\bm X}(t)) \Sigma_{0,s,r}^{<,\mathrm{st.}}(t_1,t;{\bm X}(t))  \nonumber \\
	& \hspace{10mm} + G_s^{<,\mathrm{st.}}(t,t_1;{\bm X}(t)) \Sigma_{0,s,r}^A(t_1,t) \Bigr] , \\
	& J_r^{\mathrm{ad.}}({\bm X}(t),\dot{\bm X}(t)) = 2e \sum_s \int \! dt_1 \nonumber \\
    & \hspace{3mm} \times  \mathrm{Re} \Bigl[  G_s^{R,\mathrm{st.}}(t,t_1;{\bm X}(t)) \Sigma_{0,s,r}^{<,\mathrm{ad.}}(t_1,t;{\bm X}(t),\dot{\bm X}(t))  \nonumber \\
	& \hspace{10mm} + G_s^{R,\mathrm{ad.}}(t,t_1;{\bm X}(t),\dot{\bm X}(t)) \Sigma_{0,s,r}^{<,\mathrm{st.}}(t_1,t;{\bm X}(t)) \nonumber \\
	& \hspace{10mm} + G_s^{<,\mathrm{ad.}}(t,t_1;{\bm X}(t),\dot{\bm X}(t)) \Sigma_{0,s,r}^A(t_1,t) \Bigr] .  \label{eqn:adiabatic_charge_current_def}
\end{align}
The steady-state current $J_r^{\mathrm{st.}}({\bm X}(t))$ is now expressed by the steady-state component of the GFs and the self-energies, and is easily calculated for the fixed parameter ${\bm X}(t)$.
On the other hand, the adiabatic charge current $J_r^{\mathrm{st.}}({\bm X}(t))$ is rewritten by using Eqs.~(\ref{eqn:adiabatic_reseroir_lesser_selfenergy}), (\ref{eqn:retarded_functional_derivative_first}), and (\ref{eqn:lesser_functional_derivative_first}) in the form
\begin{align}
	J_r^{\mathrm{ad.}}({\bm X}(t),\dot{\bm X}(t)) = \sum_{\mu} \pi_{r}^{\mu}(\bm{X}(t)) \dot{X}_{\mu}(t).
\end{align} 
Here, $\pi_r^{\mu}(\bm{X}(t))$ is a Berry connection given by
\begin{align}
	&\pi_{r}^{\mu}(\bm{X}(t)) =  \int \frac{d\epsilon}{2\pi} \ \mathrm{Re} \biggl[ F_1(\epsilon;\bm{X}(t)) \frac{\partial f_r(\epsilon,{\bm X}(t))}{\partial X_{\mu}} \nonumber \\
	&\hspace{3.2cm} +F_2(\epsilon;\bm{X}(t)) \frac{\partial f_{+}(\epsilon,\bm{X}(t))}{\partial X_{\mu}} \biggr],  \label{eqn:adiabatic_berry_connection} \\
	&F_1(\epsilon;\bm{X}(t)) = -e \Gamma_r \sum_s \left. \frac{\partial G_s^{R,\mathrm{st.}}(\omega;t)}{\partial \omega}\right|_{\omega=\epsilon} ,  \label{eqn:berry_connection_F_1}\\
	&F_2(\epsilon;\bm{X}(t))
	= i e \Gamma \Gamma_r \sum_{s,s^{\prime}} \int \frac{d\omega_1 d\omega_2 }{(2\pi)^2} \left( \frac{\partial }{\partial \omega_3} - \frac{\partial }{\partial \omega_4} \right) \nonumber \\
	&\hspace{1.0cm} \times \biggl[ f_r(\omega_2,{\bm X}(t)) \gamma_{s;s^{\prime}}^{R;<,\mathrm{st.}}(\omega_1,\omega_2;\omega_3,\omega_4;t)   \nonumber \\ 
	& \hspace{1.5cm} + \frac{1}{2}\gamma_{s;s^{\prime}}^{<;<,\mathrm{st.}}(\omega_1,\omega_2;\omega_3,\omega_4;t) \biggr]_{\omega_3=\omega_4=\epsilon}, \label{eqn:berry_connection_F_2}
\end{align}
where $f_+(\epsilon,\bm{X}(t))$ is an effective Fermi distribution function defined by
\begin{align}
	f_+(\epsilon,\bm{X}(t)) = \frac{\sum_r \Gamma_r f(\epsilon,T_r(t),\mu_r(t))}{\Gamma} .
	\label{eq:defeffectivedistribution}
\end{align}
Detailed derivation is given Appendix~\ref{app:Fourie_trans_theorem}.

\subsection{Pumping charge}

The pumped charge from the reservoir $r$ in one cycle is defined as
\begin{align}
	\delta Q_r = \int_{0}^{\mathcal{T}} dt \ J_r(t) . \label{eqn:charge_accumulation}
\end{align}
Substituting Eq.~(\ref{eqn:charge_current_two_app}) into Eq.~(\ref{eqn:charge_accumulation}), the pumped charge is also approximated into two terms as
\begin{align}
	\delta Q_r \simeq \delta Q_r^{\mathrm{st.}} + \delta Q_r^{\mathrm{ad.}} .
\end{align}
Here $\delta Q_r^{\mathrm{st.}}$ is the steady state term defined as
\begin{align}
	\delta Q_r^{\mathrm{st.}} &= \int^{\mathcal{T}}_{0}\! dt \, J_{r}^{\mathrm{st.}}({\bm X}(t)) .
\end{align}
This term denotes only the summation of steady charge current of fixed parameter $\bm{X}(t)$ and contains no information about transient effect of the adiabatic process.
On the other hand, $\delta Q_r^{\mathrm{ad.}}$ is the adiabatic correction term defined as
\begin{align}
	\delta Q_r^{\mathrm{ad.}} &= \int^{\mathcal{T}}_{0} \! dt \, J_{r}^{\mathrm{ad.}}({\bm X}(t),\dot{\bm X}(t)) \nonumber \\
	&= \sum_{\mu} \int_C dX_{\mu} \ \pi_{r}^{\mu}(\bm{X}) ,
\end{align}
where $C$ is a trajectory of $\bm{X}(t)$ in the parameter space.
This term is written in a geometrical manner and contains information about delayed response effect, which is discussed in the next section.

\section{\label{sec:mechanism}Pumping mechanism}

In this section, we discuss the mechanism of the charge pumping induced by reservoir parameter driving.
We first show that the present charge pumping is induced by delayed response of the QD to the time-dependent reservoir parameters in Sec.~\ref{sec:delayed_response_effect}.
Next, we show that this delay in response of the QD can be also described in terms of dynamic AC conductance in Sec.~\ref{sec:ac_response1} and Sec.~\ref{sec:ac_response2}.

\subsection{\label{sec:delayed_response_effect}Delayed response}

Let us consider the steady-state charge current for arbitrary reservoir parameters $\bm{X}(t)$ {\it delayed} with small time $\delta t (>0)$:
\begin{align}
	&J_r^{\mathrm{delay}}(t) = J_r^{\mathrm{st.}}(\bm{X}(t-\delta t)) \nonumber \\
   & \hspace{10mm} \simeq J_r^{\mathrm{st.}}(\bm{X}(t)) - \sum_{\mu} \partial^{\mu} J_r^{\mathrm{st.}}(\bm{X}(t)) \dot{X}_{\mu}(t) \delta t .
\end{align} 
It is easy to show that if we take the delay time as
\begin{align}
	\delta t = -\frac{\sum_{\mu} \pi_{r}^{\mu}(\bm{X}(t)) \dot{X}_{\mu}(t) }{\sum_{\mu} \partial^{\mu} J_r^{\mathrm{st.}}(\bm{X}(t)) \dot{X}_{\mu}(t) } , \label{eqn:delay_time_formula}
\end{align}
the correction of the steady-state current due to the time delay $\delta t$ coincides with the adiabatic correction of the current:
\begin{align}
	J_r^{\mathrm{delay}}(t) & \simeq J_r^{\mathrm{st.}}(\bm{X}(t)) + \sum_{\mu} \pi^\mu_r({\bm X}(t)) \dot{X}_{\mu}(t) \nonumber \\
    &= J_r^{\mathrm{st.}}(\bm{X}(t)) + J_r^{\mathrm{ad.}}(\bm{X}(t),\dot{\bm X}(t))
\end{align} 
We note that the definition of the delay time given in Eq.~(\ref{eqn:delay_time_formula}) holds for arbitrary strength of $U$ and $\Gamma$.
This indicates that the transient effect in the adiabatic process is always represented only by the delay time $\delta t$.
In our previous work~\cite{Hasegawa17}, we discussed this delayed response effect on the charge pumping within the first order perturbation with respect to the Coulomb interaction $U$.


\subsection{AC response: the single-reservoir case \label{sec:ac_response1}}

Before we discuss dynamic AC conductance in the present system, we consider the single-reservoir case, for which the low-frequency AC transport has been studied well~\cite{Buttiker93c,Buttiker93b,Nigg06,Lim13}.
We show that the time-dependent current under parameter driving is understood in terms of a delay time, which can be related directly to circuit elements called as dynamic capacitances and dynamic resistances.

We consider the QD coupled to one reservoir, whose temperature and electrochemical potential is modulated as
\begin{align}
	T(t) = T_0 + \delta T e^{-i \Omega t} , \quad
	\mu(t) = \mu_0 + \delta \mu \, e^{-i \Omega t} , \label{eq:paramonelead}
\end{align}
where $T_0$ and $\mu_0$ are a temperature and a electrochemical potential of the reservoir in equilibrium, and $\delta T$ and $\delta \mu$ are amplitudes of AC driving for the reservoir parameter with a frequency $\Omega$.
For convenience of description, we define the parameter vector as
\begin{align}
{\bm X}(t) &= (X_1, X_2) = \bigl(T(t),\mu(t)\bigr),
\end{align}
and rewrite Eq.~(\ref{eq:paramonelead}) as 
\begin{align}
X_\mu(t) = X_{\mu,0} + \delta X_{\mu} e^{-i\Omega t}, \quad (\mu = 1,2).
\end{align}
Here, we assume that the AC amplitude $\delta X_\mu$ is small, and consider the current flowing into the QD up to the linear order of $\delta X_\mu$:
\begin{align}
J(t) =  \sum_{\mu} G^{\mu}_{r}(\Omega;\bm{X}_0) \delta X_{\mu} e^{-i\Omega t} + O((\delta X_{\mu})^2),
   \label{eqn:def_dyn_cond}
\end{align}
where $G^{\mu}_{r}(\Omega;\bm{X}_0)$ are dynamic AC conductance.
We note that the zeroth-order term with respect to $\delta X_\mu$ is zero because the system is in thermal equilibrium for $\delta X_\mu=0$.
In the low-frequency limit, the dynamic conductance can be expanded with respect to frequency $\Omega$ as,
\begin{align}
	G^{\mu} (\Omega;\bm{X}_0) = -i \Omega G^{\mu}_1 (\bm{X}_0) - \Omega^2 G^{\mu}_2 (\bm{X}_0) 
	+ O(\Omega^3) . \label{eqn:dynamic_conductance_expand_single}
\end{align}
These coefficients, $G^{\mu}_1$ and $G^{\mu}_2$, are described by circuit elements as follows:
\begin{align}
	& G^\mu_1 = C_\mu, \ \ \quad G^\mu_2 = -C_\mu^2 R_\mu, \label{eqn:circuit_element_single}
\end{align}
where $C_\mu$ and $R_\mu$ are a dynamic capacitance and a dynamic resistance for electrochemical-potential modulation ($\mu=1$) or temperature modulation ($\mu=2$), respectively~\cite{Buttiker93c,Buttiker93b,Lim13}.
Substituting Eqs.~(\ref{eqn:dynamic_conductance_expand_single}) and (\ref{eqn:circuit_element_single}) into Eq. (\ref{eqn:def_dyn_cond}), we obtain 
\begin{align}
	J(t) &=  -i\Omega \sum_{\mu} (G^{\mu}_1 -i\Omega G^{\mu}_2)\delta X_{\mu} e^{-i\Omega t} \nonumber \\
    &\hspace{15mm} + O((\delta X_{\mu})^2,\Omega^3). \label{eq:Jdynamic}
\end{align}
This current response can be represented only by one parameter, i.e., a delay time $\delta t$ as
\begin{align}
J(t) &= J_0(t-\delta t), \label{eq:J0delay} \\
J_0(t) &=  -i\Omega \sum_{\mu} G^{\mu}_1 \delta X_{\mu} e^{-i\Omega t}, \label{eq:J0}
\end{align}
where $J_0(t)$ describe a capacitive current component due to instant response of the charge in the QD to the external parameter driving.
By comparing Eqs.~(\ref{eq:J0delay})-(\ref{eq:J0}) with Eq.~(\ref{eq:Jdynamic}), the time delay should be taken as
\begin{align}
\delta t &= -\frac{\sum_{\mu} G^{\mu,2} \delta X_{\mu} }{\sum_{\mu} G^{\mu,1} \delta X_{\mu} } =  \frac{\sum_{\mu} C_{\mu}^2 R_{\mu} \delta X_{\mu} }{\sum_{\mu} C_{\mu} \delta X_{\mu} }.
\end{align}
One can see that $\delta t$ is just an average of the relaxation time $R_{\mu}C_{\mu}$ of quantum RC circuit weighted by $C_\mu \delta X_\mu$.
This relation shows that the time delay $\delta t$ is closely related to transport coefficients in dynamic AC response of the QD.

\subsection{AC response: the two-reservoir case \label{sec:ac_response2}}

For the present system, i.e., the QD coupled to the two reservoirs, the simple interpretation by circuit elements described in the previous section is not applicable because the steady-state current generally exists.
However, we show that there is still a relation between dynamic AC response and the delay time.

We first define the parameter vector by Eq.~(\ref{eq:defXvector}), and consider the time-dependent parameter modulation give by
\begin{align}
	X_{\mu}(t) = X_{\mu,0} + \delta X_{\mu}(t) e^{-i\Omega t} .
\end{align}
The time-dependent current induced by this parameter modulation is described by
\begin{align}
	J_r(t) &= J_r^{{\rm st.}}({\bm X}_0) + 
     \sum_{\mu} G^{\mu}_{r}(\Omega;\bm{X}_0) \delta X_{\mu} e^{-i\Omega t} \nonumber \\
     & \hspace{15mm} + O((\delta X_{\mu})^2) , \label{eqnmain:def_dyn_cond}
\end{align}
where $J_r^{{\rm st.}}(t)$ is a steady-state current for a fixed parameter ${\bm X} = {\bm X}_0$, and $G^{\mu}_{r}(\Omega;\bm{X}_0)$ is a dynamic AC conductance at $\bm{X} = \bm{X}_0$.
We expand $G_r^{\mu}(\Omega;\bm{X}_0)$ with respect to $\Omega$ as
\begin{align}
	G_r^{\mu}(\Omega;\bm{X}_0) = G_{r,0}^{\mu}(\bm{X}_0) - i \Omega G_{r,1}^{\mu}(\bm{X}_0) +  O(\Omega^2) .
    \label{eqnmain:def_dyn_cond2}
\end{align}
Here, we can prove that the coefficients, $G_{r,0}^{\mu}(\bm{X}_0)$ and $G_{r,1}^{\mu}(\bm{X}_0)$, are related to the stationary current and the Berry connection as
\begin{align}
	G_{r,0}^{\mu}(\bm{X}_0) &= \partial^{\mu} J_r^{\mathrm{st.}}(\bm{X}_0), \label{eqnmain:dyn_cond_result_leading} \\
	G_{r,1}^{\mu}(\bm{X}_0) &= \pi_r^{\mu}(\bm{X}_0) , \label{eqnmain:dyn_cond_result_next_leading}
\end{align}
respectively.
Detailed derivation is given in Appendix~\ref{app:timedelay}.
This correspondence between the dynamic conductance and the Berry connection is reasonable because dynamics of the system under the low-frequency modulation is indeed an adiabatic process.

From this correspondence, we can introduce the delay time of the current under the parameter modulation, and can relate it to the low-frequency response coefficients.
Using Eqs.~(\ref{eqnmain:def_dyn_cond})-(\ref{eqnmain:dyn_cond_result_next_leading}), we obtain
\begin{align}
	J_r(t) &\simeq  J_r^{{\rm st.}}({\bm X}_0) + \delta J_r(t-\delta t), \\
    \delta J_r(t) &= \sum_{\mu} G_r^{\mu,0}(\bm{X}_0) \delta X_{\mu} e^{-i\Omega t} ,
\end{align}
where $\delta J_r(t)$ is a time-dependent current component, which instantly responses to the parameter modulation, and the time delay $\delta t$ is determined by
\begin{align}
	\delta t = - \frac{\sum_{\mu} G_r^{\mu,1}(\bm{X}_0) \delta X_{\mu} }{\sum_{\mu} G_r^{\mu,0}(\bm{X}_0) \delta X_{\mu} } . \label{eqnmain:delaytime_in_AC}
\end{align}
This expression of the delay time, which coincides with the one defined in Eq.~(\ref{eqn:delay_time_formula}), indicates that the physical picture of the delay time for the charge pumping as discussed in Sec.~\ref{sec:delayed_response_effect} is reasonable, because it is written in terms of the linear AC response to small and slow parameter modulation.

\section{Evaluation of the Pumping Charge\label{sec:application}}

In this section, we actually calculate the pumped charge by employing the renormalized perturbation theory (RPT).
We consider an approximation, the first-order perturbation with respect to the renormalized Coulomb interaction in the framework of the RPT.
After we describe the RPT (Sec.~\ref{sec:RPT}), we calculate the pumped charge as a function of $U$ and $\epsilon_d$ for the electrochemical-potential-driven charge pumping (Sec.~\ref{sec:electrochemicalpump}) and the temperature-driven one (Sec.~\ref{sec:temperaturepump}).
For simplicity, we assume the symmetric coupling, $\Gamma_L = \Gamma_R = \Gamma / 2$, and consider a symmetrized adiabatic pumped charge $\delta Q$ defined as
\begin{align}
	\delta Q &= \frac{1}{2} ( \delta Q^{\mathrm{ad.}}_L - \delta Q^{\mathrm{ad.}}_R ) \nonumber \\
   &=\frac{1}{2} \int_{C} dX_{\mu} (\pi_L^{\mu}(\bm{X}) - \pi_R^{\mu}(\bm{X}) ) .
   \label{eq:chargecountour}
\end{align}

\subsection{The renormalized perturbation theory (RPT)}
\label{sec:RPT}

In the framework of the RPT~\cite{Hewson93,Hewson01}, the original model parameters in the Anderson impurity model are replaced into the renormalized ones as
\begin{align}
\Gamma \rightarrow \tilde{\Gamma} = z \Gamma, \quad
\epsilon_d \rightarrow \tilde{\epsilon}_d, \quad
U \rightarrow \tilde{U},
\end{align}
where $z$ is a renormalization factor. 
These new parameters reflect strong renormalization due to the many-body effect.
Details of the parameter determination are given in Appendix~\ref{app:RenormalizedParameters}.

\begin{figure}[tb]
\begin{center}
\includegraphics[clip,width=7cm]{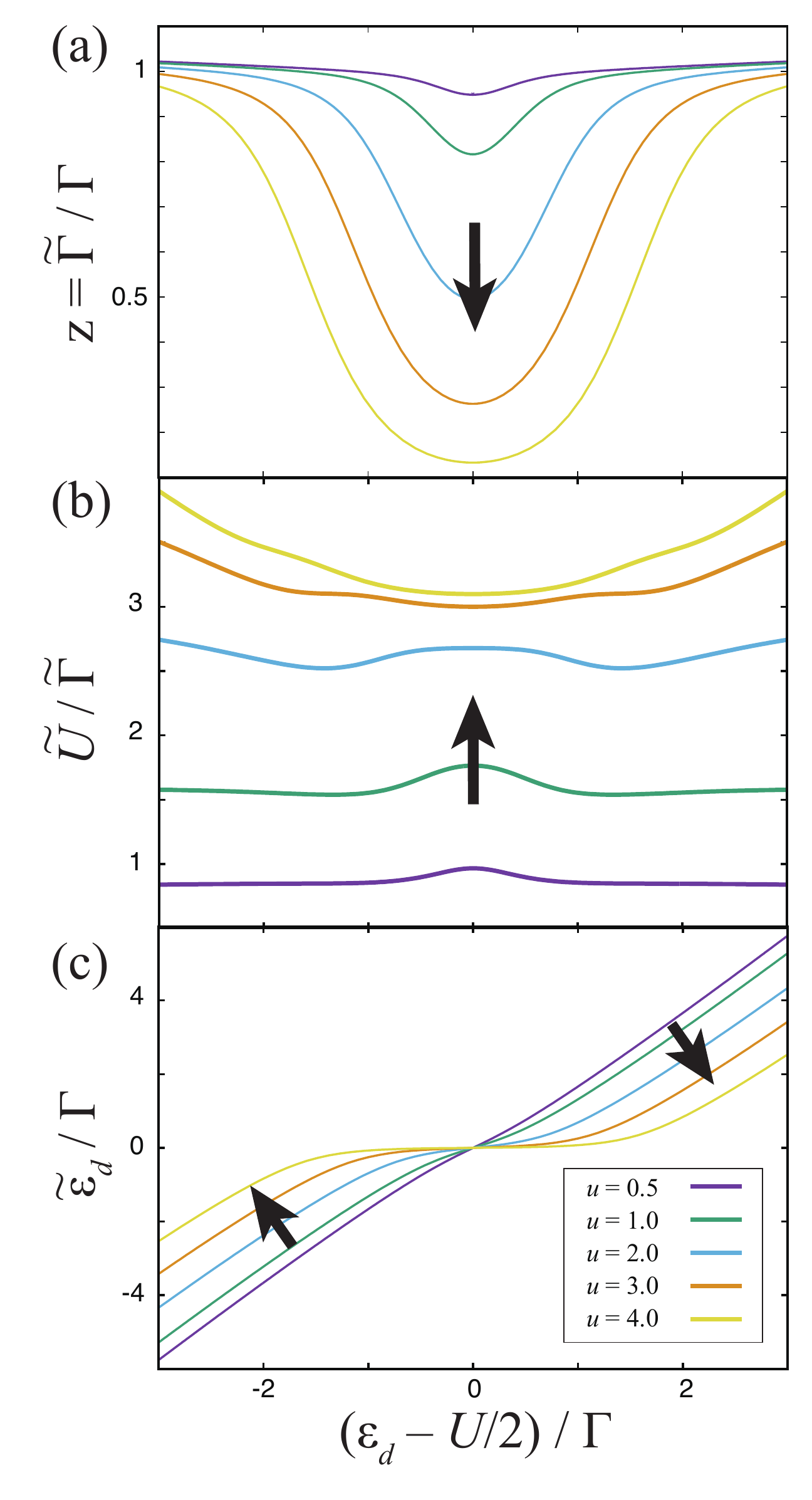}
\caption{\label{fig:renormalized_params}
The renormalized parameters are plotted as a function of $(\epsilon_d-U/2)/\Gamma$
for $u = U/\Gamma = 0.5, 1.0, 2.0, 3.0, 4.0$.
(a) The renormalized linewidth $\tilde{\Gamma}$, (b) the ratio between the renormalized Coulomb interaction $\tilde{U}$ and the renormalized line width $\tilde{\Gamma}$, and (c) the renormalized QD energy level $\tilde{\epsilon}_d$.
}
\end{center}
\end{figure}

Fig.~\ref{fig:renormalized_params} show the renormalized parameters determined from the Bethe Ansatz solution~\cite{Okiji82,Wiegmann83} for several values of $u \equiv U/\Gamma$.
The effective linewidth $\tilde{\Gamma}$ indicates the peak width of the density of states for the QD, and gives the characteristic energy scale of the system. 
In the presence of the Coulomb interaction, $\tilde{\Gamma}$ is strongly suppressed due to the many-body effect inherent to the Kondo problem.
Actually, as seen in Fig.~\ref{fig:renormalized_params}~(a), the renormalized linewidth $\tilde{\Gamma}$ is reduced in the presence of the Coulomb interaction when the occupation of the QD is near one.
The Coulomb interaction $\tilde{U}$ is also renormalized as shown in Fig.~\ref{fig:renormalized_params}~(b);
the ratio $\tilde{U}/\tilde{\Gamma}$ first increases as $u$ increases, and shows a tendency of saturation for $u \gtrsim 3$.
The renormalized QD energy level is shown in Fig.~\ref{fig:renormalized_params}~(c); it becomes flat near the particle-hole symmetric point due to the pinning effect because the occupation number of the electron in the QD is fixed almost at one for the strong Coulomb interaction.

In the following calculation, we consider the first-order perturbation theory for the renormalized model.
In our previous paper~\cite{Hasegawa17}, we have calculated the pumped charge up to the first-order perturbation for the bare model parameters.
The present result is obtained by replacing the model parameters in Ref.~\onlinecite{Hasegawa17} with the renormalized ones.

We give some remarks on the limitation of the first-order perturbation based on the RPT.
First, because the RPT is based on the Fermi liquid theory, it is applicable when the temperature is sufficiently small compared with the renormalized linewidth $\tilde{\Gamma}$, i.e., the Kondo temperature.
Next, since we employ the first order perturbation theory with respect to the renormalized Coulomb interaction $\tilde{U}$, the present results ignore the higher order effects.
In this paper, however, we focus on the qualitative tendency of how the parameter renormalization by the Coulomb interaction modifies the pumped charge.
For this purpose, the first-order perturbation is sufficient, because its major effect of parameter renormalization is described in the present approximation.

\subsection{Electrochemical-potential-driven pumping}
\label{sec:electrochemicalpump}

First, let us consider electrochemical-potential-driven pumping.
We set the temperatures of the reservoirs as zero, and consider only the electrochemical-potential driving in the near-equilibrium region,
\begin{align}
	\mu_r(t) &= \epsilon_F + \delta \mu_r (t),
\end{align}
where $\epsilon_F$ is the Fermi level (set as zero throughout this paper), and $\delta \mu_r (t)$ is a time-dependent part of the electrochemical potential of the reservoir $r$.
We assume that the amplitude of the time-dependent part is small:
\begin{align}
	\delta \mu \ll \tilde{\Gamma} < \Gamma,
\end{align}
where $\delta \mu= {\rm max} |\mu(t)|$.
By the Stokes theorem, the symmetrized adiabatic pumped charge given in Eq.~(\ref{eq:chargecountour}) is rewritten as
\begin{align}
	\delta Q &= \frac{1}{2} \int_{A} d\mu_L d\mu_R \biggl[ \frac{\partial (\pi_L^{4}(\bm{X}) - \pi_R^{4}(\bm{X}))}{\partial \mu_L} \nonumber \\
	& \hspace{2.5cm} - \frac{\partial (\pi_L^{3}(\bm{X}) - \pi_R^{3}(\bm{X}))}{\partial \mu_R} \biggr] , \label{eqn;sym_pump_charge_mu}
\end{align}
where $A$ indicates an integral surface on the $\mu_L$-$\mu_R$ plane whose boundary is $C$.
For the small-amplitude driving of electrochemical potentials ($\delta \mu \ll \tilde{\Gamma}$), $\delta Q$ can be approximated as
\begin{align}
	& \delta Q = \Pi_{0,\rm{volt.}} V(A) + O((\delta \mu/\Gamma)^3), \\
   	& V(A) = \Gamma^{-2} \int_A d\mu_L d\mu_R, 
\end{align}
where $V(A)$ is a dimensionless quantity proportional to the area inside the contour $C$ in the $\mu_L$-$\mu_R$ plane.
The kernel $\Pi_{0,\rm{volt.}}$, which indicates strength of the pumping, is calculated at zero temperature as
\begin{align}
	& \Pi_{0,\rm{volt.}} = \frac{e \Gamma^4}{8\pi^2} \mathrm{Im} \gamma_{0}(0,0), \label{eqn:def_of_large_pi_volt} \\
	& \gamma_0(\omega,\epsilon) = \int \frac{d\omega_1}{2\pi} \sum_{s,s^{\prime}} \left(-\frac{\partial}{\partial \omega_1} -\frac{\partial}{\partial \omega_2} - \frac{\partial}{\partial \omega_3} + \frac{\partial}{\partial \omega_4} \right) \nonumber \\
	&\hspace{1.0cm} \times \left. \gamma_{s;s^{\prime}}^{R;<,\mathrm{st.}}(\omega_1,\omega_2;\omega_3,\omega_4) \right|_{\omega_2=\omega, \omega_3=\omega_4 = \epsilon} . \label{eqn:def_of_gamma_zero} 
\end{align}
Now, we apply the first-order perturbation in the framework of the RPT to Eqs. (\ref{eqn:def_of_large_pi_volt}) and (\ref{eqn:def_of_gamma_zero}).
The four-point vertex function $ \gamma_{s;s^{\prime}}^{R;<,\mathrm{st.}}$ is written into renormalized GFs as 
\begin{align}
	&\gamma_{s;s^{\prime}}^{R;<,\mathrm{st.}}(\omega_1,\omega_2;\omega_3,\omega_4) \nonumber \\
	&\hspace{5mm} = -i z^2 \tilde{U} \tilde{G}_{s,0}^R(\omega_1) \tilde{G}_{s,0}^R(\omega_2) \tilde{G}_{s^{\prime},0}^R(\omega_3) \tilde{G}_{s^{\prime},0}^A(\omega_4) ,
\end{align}
where $\tilde{G}_{s,0}^{R}(\omega)$ and $\tilde{G}_{s,0}^{A}(\omega)$ is a retarded and advanced renormalized GF, respectively, defined as
\begin{align}
	& \tilde{G}_{s,0}^{R}(\omega) = \frac{1}{\omega - \tilde{\epsilon}_d + i \tilde{\Gamma} / 2} , \\
   & \tilde{G}_{s,0}^{A}(\omega) = \frac{1}{\omega - \tilde{\epsilon}_d - i \tilde{\Gamma} / 2} .
\end{align}
As a result, $\Pi_{0,\rm{volt.}}$ is written as
\begin{align}
	\Pi_{0,\rm{volt.}} &= - \frac{e}{8\pi^2} \frac{ z^2 \tilde{U} \tilde{\epsilon}_d \tilde{\Gamma}^6}{(\tilde{\epsilon}_d^2 + \tilde{\Gamma}^2 /4 )^4} + O(\tilde{U}^2) . \label{eqn:pi_0_mu_formula} 
\end{align}

\begin{figure}[tb]
\begin{center}
\includegraphics[clip,width=9cm]{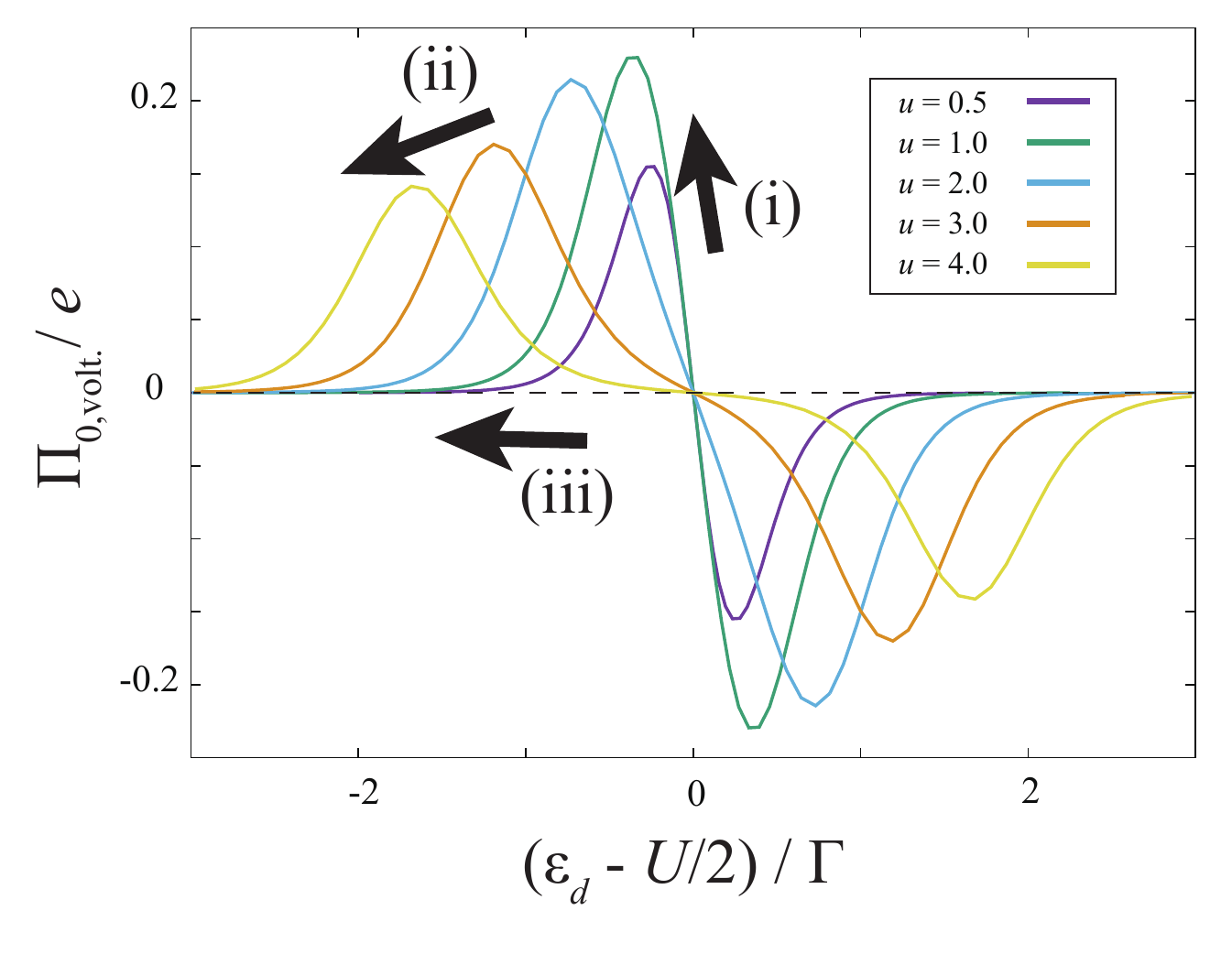}
\caption{\label{fig:Uplot_mu}
The QD energy level dependence of $\Pi_{0,\mathrm{volt.}}$ for different values of the Coulomb interaction.
Five results are plotted for $u = U/\Gamma = 0.5, 1.0, 2.0, 3.0, 4.0$.
As the Coulomb interaction becomes stronger, the amplitude is enhanced for $u \lesssim 1.0$ and suppressed for  $u \gtrsim 1.0$.
The shape gets more and more broadened and the peak and dip position is shifted to larger value.
}
\end{center}
\end{figure}

In Fig.~\ref{fig:Uplot_mu}, we plot $\Pi_{0,\rm{volt.}}$ as a function of $\epsilon_d$ for several values of $U$.
As seen from the figure, $\Pi_{0,\rm{volt.}}$ is an odd function with respect to $\epsilon_d-U/2$, which is a deviation from the particle-hole symmetric point, and changes its sign at $\epsilon_d = U/2$.
The qualitative change of $\Pi_{0,\rm{volt.}}$ for increasing the Coulomb interaction is summarized as follows; (i) the amplitude of its peak is at first enhanced for $U \lesssim \Gamma$ due to the increase of $\tilde{U}/\tilde{\Gamma}$, (ii) it is suppressed when the Coulomb interaction $U$ proceeds $\Gamma$ because the suppression of the renormalization linewidth $\tilde{\Gamma}$ becomes relevant, and (iii) the peak position moves away from the particle-hole symmetric point because of the pinning effect for the renormalized QD energy level $\tilde{\epsilon}_d$.

\subsection{Temperature-driven pumping}
\label{sec:temperaturepump}

Next, let us consider the temperature-driven pumping.
We set the electrochemical potentials to the Fermi energy (set as zero throughout this paper), and consider only the temperature driving in the near-equilibrium region,
\begin{align}
	T_r(t) &= T_0 + \delta T_r (t) ,
\end{align}
where $T_0$ is an average temperature, and $\delta T_r(t)$ is a time-dependent part of the temperature of the reservoir $r$.
We assume that the amplitude of the time-dependent part is small:
\begin{align}
	\delta T \ll T_0 \ll \tilde{\Gamma} < \Gamma , 
\end{align}
where $\delta T= \mathrm{max} |\delta T_r(t)|$.

In the same manner as the electrochemical potential driving, the symmetrized adiabatic pumped charge is written as 
\begin{align}
	& \delta Q = \Pi_{0,\mathrm{temp.}} V(A) + O((\delta T/\Gamma)^3), \\
   & V(A) = \Gamma^{-2} \int_A dT_L dT_R  .
\end{align}
where $V(A)$ is a dimensionless quantity proportional to the area inside the contour $C$ in the $T_L$-$T_R$ plane.
The kernel $\Pi_{0,\rm{temp.}}$, which indicates strength of the temperature pumping, is calculated as
\begin{align}
	\Pi_{0,\mathrm{temp.}}&= \frac{e \pi^2 \Gamma^4 T_0^2}{72} \mathrm{Im} \left. \frac{\partial^2}{\partial \omega \partial \epsilon} \gamma_{0}(\omega,\epsilon) \right|_{\omega=\epsilon=0}, \label{eqn:def_of_large_pi_temp}
\end{align}
Applying the first-order perturbation approximation to $\Pi_{0,\mathrm{temp.}}$, we obtain
\begin{align}
	\Pi_{0,\mathrm{temp.}} &= - \frac{e\pi^2 z^2 T_0^2}{18} \frac{\tilde{U} \tilde{\Gamma}^6 \tilde{\epsilon}_d ( \tilde{\Gamma}^2/4 - 3 \tilde{\epsilon}_d^2 )}{(\tilde{\epsilon}_d^2 + \tilde{\Gamma}^2 / 4)^6} .  \label{eqn:pi_0_temp_formula}
\end{align}

\begin{figure}[tb]
\begin{center}
\includegraphics[clip,width=9cm]{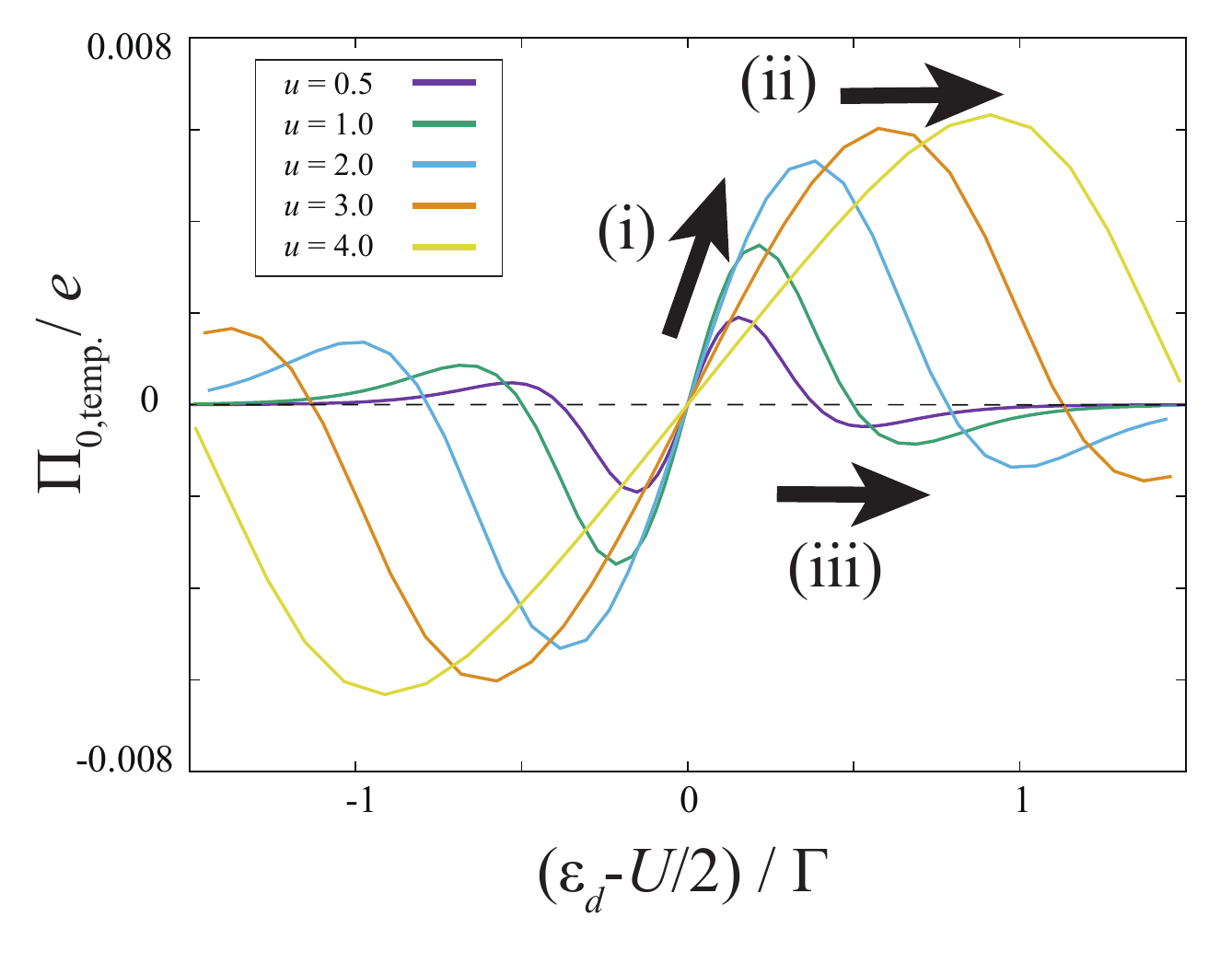}
\caption{\label{fig:Uplot_temp}
The QD energy level dependence of $\Pi_{0,\mathrm{volt.}}$ for different values of the Coulomb interaction.
Five results are plotted for $u = U/\Gamma = 0.5, 1.0, 2.0, 3.0, 4.0$.
As the Coulomb interaction becomes stronger, the amplitude is enhanced for $u \lesssim 1.0$ and suppressed for  $u \gtrsim 1.0$.
The shape gets more and more broadened and the peak and dip position is shifted to larger value.
}
\end{center}
\end{figure}

In Fig.~\ref{fig:Uplot_temp}, we plot $\Pi_{0,\rm{temp.}}$ as a function of $\epsilon_d$ for several values of $U$.
As seen from the figure, $\Pi_{0,\rm{temp.}}$ is an odd function with respect to $\epsilon_d-U/2$, and changes its sign three times at $\epsilon_d = U/2$, and other two values of $\epsilon_d$.
The qualitative change of $\Pi_{0,\rm{temp.}}$ for increasing the Coulomb interaction is summarized as follows; (i) the amplitude of its peak continues to increase due to the increase of $\tilde{U}/\tilde{\Gamma}$, (ii) its peak growth saturates due to the saturation of $\tilde{U}/\tilde{\Gamma}$ for $u \gtrsim 3$, and (iii) the peak position moves away from the particle-hole symmetric point because of the pinning effect for the renormalized QD energy level $\tilde{\epsilon}_d$.
The behavior (ii) is different from the electrochemical potential driving.
This is because $\Pi_{0,\rm{temp.}}$ only depends on $\tilde{U}/\tilde{\Gamma}$ at the peak $\tilde{\epsilon}_d \simeq 0.12 \tilde{\Gamma}$, and is independent of the renormalization factor $z$ (see Eq.~(\ref{eqn:pi_0_temp_formula})).

\section{Summary \label{sec:summary}}

In this paper, we have studied adiabatic charge pumping via a single-level QD system induced by reservoir parameter driving in the coherent transport region.
Based on the Anderson impurity model, we have formulated the pumped charge in the presence of the Coulomb interaction by the Keldysh formalism.
To handle the time-dependent reservoir temperatures, we have introduced a thermomechanical field, which describes temperature modulation via an energy rescaling of the reservoirs.
Applying the adiabatic approximation, we have derived the formula for the adiabatic pumped charge in terms of the Berry connection, and have shown that it is expressed by the two-particle Green's function of electrons in the QD.
Our formalism covers the low-temperature and strongly-correlated region, which cannot be treated in the previous theoretical methods.

The origin of the present adiabatic charge pumping is a delayed response of the QD with respect to the reservoir parameter driving.
We have shown that the Berry connection is generally expressed in terms of the time delay, and that this time delay coincides with the characteristic relaxation time determined from the dynamic AC response.
While our discussion is analogous to that of the relaxation time in quantum RC circuits, our discussion is applicable also to the nonequilibrium states in the interacting electron systems under a finite bias.

Based on our formulation, we have calculated the pumped charge by employing the first-order perturbation in the framework of the renormalized perturbation theory (RPT).
We have shown how the parameter renormalization in the RPT affects the adiabatic charge pumping, for the electrochemical-potential-driven and the temperature-driven case, as the Coulomb interaction becomes larger.
For both of the cases, the pumping strength has a maximum and a minimum as a function of the QD energy level, and the position of the maximum (minimum) moves away from the particle-hole symmetric point as the Coulomb interaction increases.
For the electrochemical-potential-driven case, the maximum value of the pumping strength is once enhanced, and then suppressed as the Coulomb interaction increases.
On the other hand, for the temperature-driven case, the maximum value continues to increases as the Coulomb interaction increases.
This difference has been explained from the renormalized parameters consistently.

Our main result is the formulation of adiabatic charge pumping induced by reservoir parameter driving, which is applicable to the low temperature and strong Coulomb interaction region.
Our formalism states that adiabatic charge pumping can be evaluated by the two-particle Green's function of the electrons in the QD, which is in principle calculated by numerical methods, such as numerical renormalization group and continuous-time quantum Monte Carlo method.
It is an important future problem to compute the pumped charge accurately in the strong Coulomb interaction region.
From the viewpoint of nonequilibrium thermodynamics, it is also a challenge to generalize the present formalism to heat pumping and work exchange and to discuss quantum effects on efficiency of small engines.

\begin{acknowledgments}
The authors greatly thank R. Sakano for helping the numerical calculation.
Also the authors thank T. Yokoyama, H. Shinaoka for helpful discussions.
This work was supported by Japan Society for the Promotion of Science KAKENHI Grants No. JP24540316 and No. JP26220711.
M.H. acknowledges financial support provided by the Advanced Leading Graduate Course for Photon Science.
\end{acknowledgments}

\appendix

\section{\label{app:adiabtic_selfenergy} Reservoir self-energies}

In this appendix, we calculate the reservoir self-energies, and derive Eqs.~(\ref{eqn:retarded_reservoir_selfenergy})-(\ref{eqn:adiabatic_reseroir_lesser_selfenergy}).
The retarded component of the reservoir self-energy is calculated as follows:
\begin{align}
	& \Sigma_{0,s,r}^R(t_1,t_2) = \sum_{k,s} v_r^{\ast}(t_1) g^R_{rks}(t_1,t_2) v_r(t_2) \nonumber \\
	& \hspace{3mm} = -i \int \! \frac{d\epsilon}{2\pi} \Gamma_r \sqrt{B_r(t_1) B_r(t_2)} \Theta (t_1-t_2) e^{-i(I_1 +I_2 \epsilon)} \nonumber \\
	& \hspace{3mm} = -i \frac{\Gamma_r}{2} \delta(t_1-t_2), \label{eqn:retarded_selfenergy_derivation} 
\end{align}
Here $\Theta(t_1-t_2)$ is the Heaviside step function, $I_1 = \int_{t_2}^{t_1} dt \, \mu_r(t)$, and $I_2= \int_{t_2}^{t_1} dt \, B_r(t)$.
In the last line of Eq.~(\ref{eqn:retarded_selfenergy_derivation}), we have used a formula for the delta function:
\begin{align}
	\delta \left( \int_{t_2}^{t_1}dt \ B_r(t) \right) = \frac{1}{B_r(t_1)} \delta(t_1-t_2) .
\end{align}
This formula holds because $B_r(t)$ is always positive and the equation, $\int_{t_2}^{t_1}dt \ B_r(t) = 0$, has only one solution, $t_1 = t_2$.
The advanced component is obtained as a complex conjugate of the retarded one as
\begin{align}
	\Sigma_{0,s,r}^A(t_1,t_2) &= \left( \Sigma_{0,s,r}^R(t_2,t_1) \right)^{\ast} 
	= i \frac{\Gamma_r}{2} \delta(t_1-t_2) . \label{eqn:advanced_selfenergy_derivation}
\end{align}
As seen in Eqs.~(\ref{eqn:retarded_selfenergy_derivation}) and (\ref{eqn:advanced_selfenergy_derivation}), in the wide-band limit, the retarded and advanced components have no dependence of reservoir parameters and so do not have any adiabatic correction term.

The lesser component is calculated as follows:
\begin{align}
	& \Sigma_{0,s,r}^<(t_1,t_2) = \sum_{k,s} v_r^{\ast}(t_1) g^<_{rks}(t_1,t_2) v_r(t_2) \nonumber \\
	& \hspace{3mm} = i \int \! \frac{d\epsilon}{2\pi} \Gamma_r \sqrt{B_r(t_1) B_r(t_2)} f(\epsilon) e^{-i(I_1 + \epsilon I_2)}.
    \label{eqn:self_energy_lesser_general}
\end{align}
Using the adiabatic approximation given in Eqs.~(\ref{eqn:add_app_tmf}) and (\ref{eqn:add_app_mu}), and ignoring higher order terms such as $\ddot{B}_r(t)$ and $(\dot{B}_r(t))^2$, we obtain 
\begin{align}
& \sqrt{\frac{B_r(t_1)}{B_r(t_2)}} \simeq 1 + \frac{1}{2} \frac{\dot{B}_r(t)}{B_r(t)} \bigl[ (t_1 - t) - (t_2 - t) \bigr], \\
& \hspace{5mm} I_1 \simeq \mu_r(t) (t_1 - t_2) \nonumber \\
& \hspace{12mm} + \frac12 \dot{\mu}_r(t) \bigl[ (t_1 - t)^2 - (t_2 - t)^2 \bigr],  \\
& \hspace{5mm} I_2 \simeq B_r(t) (t_1 - t_2) \nonumber \\
& \hspace{12mm} + \frac 12 \dot{B}_r(t) \bigl[ (t_1 - t)^2 - (t_2 - t)^2 \bigr]. 
\end{align}
Substituting these expressions, and ignoring higher order terms again, we obtain
\begin{align}
	&\Sigma_{0,s,r}^<(t_1,t_2)  \nonumber \\
	&= i \Gamma_r \left\{ 1+\frac{1}{2}\frac{\dot{T}_r(t)}{T_r(t)} \bigl[(t_1-t) + (t_2-t)\bigr]  \right\} \nonumber \\
	& \hspace{0.5cm} \times \int \frac{d\epsilon}{2\pi}  f(\epsilon, T_r(t),\mu_r(t)) e^{-i \epsilon (t_1-t_2)} \nonumber \\
	& \hspace{0.5cm} \times \biggl\{ 1- \frac{i}{2} \left( \frac{\dot{T}_r(t)}{T_r(t)} ( \epsilon -\mu_r(t)) + \dot{\mu}_r(t) \right) \nonumber \\
	& \hspace{1cm} \times \bigl[(t_1-t)^2 - (t_2-t)^2 \bigr] \biggr\}  . \label{eqn:add_app_lesser_derivation}
\end{align}
Here, the integral variable is converted from $\epsilon$ to $\epsilon^\prime = (\epsilon-\mu_r(t))/B_r(t)$, and set $\epsilon^\prime$ to $\epsilon$ again.

\section{\label{app:derivation_ward_identity}Functional derivatives}

In this appendix, we derive explicit expressions of the functional derivatives, Eqs.~(\ref{eqn:funcder_GF_to_4point_ret}) and (\ref{eqn:funcder_GF_to_4point_les}).
We also show that the functional derivatives are related to the two-particle GFs.
Our derivation is based on a standard method, which is seen in the textbook of the field theory, and does not depend a specific approximation such as the adiabatic approximation.
For simplicity, we take time variables on the Keldysh contour, and project them onto real-time contour at the end of this appendix.

We first decompose the functional derivative into two factors as
\begin{align}
	&\frac{\delta G_{s_1}(\tau_1,\tau_2)}{\delta \Sigma_{0,s_2,r}(\tau_4,\tau_3)} \nonumber \\
    & \hspace{5mm} = \int d\tau_5 d\tau_6 \ \frac{\delta G_{s_1}(\tau_1,\tau_2)}{\delta \Sigma_{s_1}(\tau_5,\tau_6)} \frac{\delta \Sigma_{s_1}(\tau_5,\tau_6)}{\delta \Sigma_{0,s_2,r}(\tau_4,\tau_3)} , \label{eqn:Ward_G/Sigma_0}
\end{align}
By Dyson's equation, the former factor is rewritten as
\begin{align}
	&\frac{\delta G_{s_1}(\tau_1,\tau_2)}{\delta \Sigma_{s_1}(\tau_3,\tau_4)}  = G_{s_1}(\tau_1,\tau_3) G_{s_1}(\tau_4,\tau_2) ,
\end{align}
while the latter factor is calculated as
\begin{align}
	\frac{\delta \Sigma_{s_1}(\tau_1,\tau_2)}{\delta \Sigma_{0,s_2}(\tau_4,\tau_3)} &= \delta_{s_1,s_2} \delta(\tau_1-\tau_4) \delta(\tau_2-\tau_3) \nonumber \\
	& \hspace{1.0cm} + \frac{\delta \Sigma_{U,s_1}(\tau_1,\tau_2)}{\delta \Sigma_{0,s_2}(\tau_4,\tau_3)} .\label{eqn:Ward_Sigma_U/Sigma_0}
\end{align}
The functional derivative of the interaction self-energy is rewritten into 
\begin{align}
	&\frac{\delta \Sigma_{U,s_1}(\tau_1,\tau_2)}{\delta \Sigma_{0,s_2}(\tau_4,\tau_3)}  \nonumber \\
	&= \sum_{s_3} \int d\tau_5 d\tau_6 \ I_{s_1;s_3} (\tau_1,\tau_2;\tau_5,\tau_6) \frac{\delta G_{s_3}(\tau_6,\tau_5)}{\delta \Sigma_{0,s_2}(\tau_4,\tau_3)} , \label{eqn:Ward_identity}
\end{align}
where $I_{s_1;s_3} (\tau_1,\tau_2;\tau_5,\tau_6)$ is a two-particle irreducible (2PI) four-point vertex function.
Substituting Eqs.~(\ref{eqn:Ward_G/Sigma_0})-(\ref{eqn:Ward_Sigma_U/Sigma_0}) into Eq.~(\ref{eqn:Ward_identity}), it becomes
\begin{align}
	&\sum_{s_3} \int \! d\tau_5 d\tau_6 \, R_{s_1;s_3}(\tau_1,\tau_2;\tau_6,\tau_5) \frac{\delta \Sigma_{U,s_3}(\tau_5,\tau_6)}{\delta \Sigma_{0,s_2}(\tau_4,\tau_3)}  \nonumber \\
	&\hspace{5mm} = \sum_{s_3} \int \! d\tau_5 d\tau_6 \, I_{s_1;s_3} (\tau_1,\tau_2;\tau_5,\tau_6) \nonumber \\
    & \hspace{30mm} \times G_{s_3}(\tau_3,\tau_5) G_{s_3}(\tau_6,\tau_4)  , \label{eqn:smilar_bethe_salpeter}
\end{align}
where
\begin{align}
	&R_{s_1;s_3}(\tau_1,\tau_2;\tau_6,\tau_5) = \delta_{s_1,s_3} \delta(\tau_1-\tau_5)  \delta(\tau_2-\tau_6) \nonumber \\
	&\hspace{0.3cm} - \int \! d\tau_7 d\tau_8 \, I_{s_1;s_3}(\tau_1,\tau_2;\tau_7,\tau_8) G_{s_3}(\tau_6,\tau_7) G_{s_3}(\tau_8,\tau_5) .
\end{align}
The formal solution of Eq.~(\ref{eqn:smilar_bethe_salpeter}) is written as
\begin{align}
	&\frac{\delta \Sigma_{U,s_1}(\tau_1,\tau_2)}{\delta \Sigma_{0,s_2}(\tau_4,\tau_3)}  \nonumber \\
	&= \sum_{s_3,s_4} \int \! d\tau_5 \cdots d\tau_8 \, R^{-1}_{s_1;s_3}(\tau_1,\tau_2;\tau_5,\tau_6) \nonumber \\
    &\hspace{5mm} \times I_{s_3;s_4}(\tau_6,\tau_5;\tau_7,\tau_8) G_{s_3}(\tau_3,\tau_7) G_{s_3}(\tau_8,\tau_4) . \label{eqn:formal_solution}
\end{align}
Here $R^{-1}$ is an inverse function which satisfies
\begin{align}
	&\sum_{s_3 }\int \! d\tau_5 d\tau_6 \, R^{-1}_{s_1;s_3}(\tau_1,\tau_2;\tau_5,\tau_6) R_{s_3;s_2}(\tau_6,\tau_5;\tau_3,\tau_4) \nonumber \\
	&\hspace{3.0cm} = \delta_{s_1,s_2} \delta(\tau_1-\tau_3) \delta(\tau_2-\tau_4) .
\end{align}
By the Bethe-Salpeter equation, the product of $R^{-1}$ and $I$ equals to a full four-point vertex function, denoted as
\begin{align}
	& \Gamma_{s_1;s_2}(\tau_1,\tau_2;\tau_3,\tau_4) \nonumber \\
	&=\sum_{s_3} \int \! d\tau_5 d\tau_6 \, R^{-1}_{s_1;s_3}(\tau_1,\tau_2;\tau_5,\tau_6) I_{s_3;s_2}(\tau_6,\tau_5;\tau_3,\tau_4) , \label{eqn:full_vertex}
\end{align}
where $\Gamma_{s_1;s_2}(\tau_1,\tau_2;\tau_3,\tau_4)$ is a full four-point vertex function.
Substituting Eq.~(\ref{eqn:full_vertex}) into Eq. (\ref{eqn:formal_solution}) and Eqs.~(\ref{eqn:Ward_G/Sigma_0}) and (\ref{eqn:Ward_Sigma_U/Sigma_0}), we obtain
\begin{align}
	&\frac{\delta G_{s_1}(\tau_1,\tau_2)}{\delta \Sigma_{0,s_2}(\tau_4,\tau_3)} = G_{s_1}(\tau_1,\tau_4) G_{s_1}(\tau_3,\tau_2) \delta_{s_1,s_2} \nonumber \\
	& \hspace{0.3cm} +\int \! d\tau_5 \cdots d\tau_8 \, \Bigl[ G_{s_1}(\tau_1,\tau_5)G_{s_1}(\tau_6,\tau_2) \nonumber \\
	& \hspace{0.6cm} \times  \Gamma_{s_1;s_2}(\tau_5,\tau_6;\tau_7,\tau_8) G_{s_2}(\tau_3,\tau_7) G_{s_2}(\tau_8,\tau_4) \Bigr] . \label{eqn:G/Sigma_on_Keldysh}
\end{align}
By choosing appropriate branches for the time variables, this equation gives explicit expressions for the functional derivatives given in Eqs.~(\ref{eqn:funcder_GF_to_4point_ret}) and (\ref{eqn:funcder_GF_to_4point_les}).
Eq. (\ref{eqn:G/Sigma_on_Keldysh}) is rewritten in a simple form by introducing a connected two-particle GF defined by
\begin{align}
	D_{s;s^{\prime},c}(\tau_1,\tau_2;\tau_3,\tau_4) 
    &= \Braket{T_K  d_{s}(\tau_1) d_{s^{\prime}}(\tau_3) d_{s}^{\dagger}(\tau_2) d_{s^{\prime}}^{\dagger}(\tau_4)} \nonumber \\
	& - G_{s}(\tau_1,\tau_2)G_{s^{\prime}}(\tau_3,\tau_4) \nonumber \\
    & + G_{s}(\tau_1,\tau_4)G_{s}(\tau_3,\tau_2) \delta_{s,s^{\prime}} . \label{eqn:def_of_connected_4_point_correlation}
\end{align}
The functional derivative of GF is then written as
\begin{align}
	\frac{\delta G_{s_1}(\tau_1,\tau_2)}{\delta \Sigma_{0,s_2}(\tau_4,\tau_3)} &= G_{s_1}(\tau_1,\tau_4) G_{s_1}(\tau_3,\tau_2) \delta_{s_1,s_2} \nonumber \\
	&\hspace{1.0cm}+ D_{s_1;s_2,c}(\tau_1,\tau_2;\tau_3,\tau_4) .
\end{align}
Finally, we project time variables onto the real-time contour.
The functional derivative of the retarded GF becomes
\begin{align}
	\frac{\delta G_{s_1}^{R}(t_1,t_2)}{\delta \Sigma_{0,s_2}^{<}(t_4,t_3)} &= \gamma_{s_1,s_2}^{R;<}(t_1,t_2;t_3,t_4) , \nonumber \\
	&\hspace{-15mm}= \sum_{\nu_3,\nu_4} (-1)^{\nu_3}(-1)^{\nu_4} \Bigl[ D_{s_1;s_2,c}^{++\nu_3 \nu_4}(t_1,t_2;t_3,t_4) \nonumber \\
    & \hspace{5mm}- D_{s_1;s_2,c}^{+-\nu_3 \nu_4}(t_1,t_2;t_3,t_4) \Bigr] ,
\end{align}
where $\nu_3$ and $\nu_4$ are the Keldysh component index for $t_3$ and $t_4$, respectively, and $(-1)^{\nu}$ equals to 1 for $\nu=+$ and $-1$ for $\nu=-$.
The functional derivative of the lesser GF becomes
\begin{align}
	\frac{\delta G_{s_1}^{<}(t_1,t_2)}{\delta \Sigma_{0,s_2}^{<}(t_4,t_3)} &= \gamma_{s_1,s_2}^{<;<}(t_1,t_2;t_3,t_4), \nonumber \\
	&\hspace{-15mm} = G_{s_1}^R(t_1,t_4) G_{s_1}^A(t_1,t_4) \delta_{s_1,s_2} \nonumber \\
	&\hspace{-15mm} + \sum_{\nu_3,\nu_4} (-1)^{\nu_3}(-1)^{\nu_4}  D_{s_1;s_2,c}^{+-\nu_3 \nu_4}(t_1,t_2;t_3,t_4) .
\end{align}

\section{\label{app:Fourie_trans_theorem} Adiabatic pumping current}

In this appendix, we derive the expressions of the adiabatic charge current, Eqs.~(\ref{eqn:adiabatic_berry_connection})-(\ref{eqn:berry_connection_F_2}) from the definition, Eq.~(\ref{eqn:adiabatic_charge_current_def}), which is
is rewritten by using Eqs.~(\ref{eqn:retarded_functional_derivative_first})-(\ref{eqn:funcder_GF_to_4point_les}) as
\begin{align}
	&J_r^{\mathrm{ad.}}(\bm{X}(t),\dot{\bm{X}}(t))
	= 2e \sum_s \int dt_1   \nonumber \\
	&\hspace{5mm} \times \mathrm{Re} \Bigl[ G_s^{R,\mathrm{st.}}(t,t_1;\bm{X}(t)) \Sigma_{0,s,r}^{<,\mathrm{ad.}}(t_1,t;\bm{X}(t),\dot{\bm{X}}(t))  \nonumber \\
	&\hspace{5mm} + \int \! dt_2 \! \int \! dt_3 \gamma_{s;s^{\prime}}^{R;<,\mathrm{st.}}(t,t_1;t_2,t_3;\bm{X}(t)) \nonumber \\
    & \hspace{15mm} \times \Sigma_{0,s}^{<,\mathrm{ad.}}(t_3,t_2;\bm{X}(t),\dot{\bm{X}}(t)) \Sigma_{0,s,r}^{<,\mathrm{st.}}(t_1,t;\bm{X}(t)) \nonumber \\
	&\hspace{5mm} + \int \! dt_2 \! \int \! dt_3 \gamma_{s;s^{\prime}}^{<;<,\mathrm{st.}}(t,t_1;t_2,t_3;\bm{X}(t)) \nonumber \\
    & \hspace{15mm} \times \Sigma_{0,s}^{<,\mathrm{ad.}}(t_3,t_2;\bm{X}(t),\dot{\bm{X}}(t)) \Sigma_{0,s,r}^{A}(t_1,t) \Bigr] . \label{eqn:adiabatic_charge_current_time}
\end{align}
Substituting Eq.~(\ref{eqn:adiabatic_reseroir_lesser_selfenergy}), the integral of the first term in r.h.s of Eq.~(\ref{eqn:adiabatic_charge_current_time}) is rewritten as
\begin{align}
	&\int \! dt_1 G_s^{R,\mathrm{st.}}(t,t_1;\bm{X}(t)) \Sigma_{0,s,r}^{<,\mathrm{ad.}}(t_1,t;\bm{X}(t),\dot{\bm{X}}(t)) \nonumber \\
	&= \int \! dt_1 \int \! \frac{d\omega d\epsilon}{(2\pi)^2} G_s^{R,\mathrm{st.}}(\omega;\bm{X}(t))  \Gamma_r   f(\epsilon, T_r(t),\mu_r(t))   \nonumber \\
	& \hspace{0.25cm} \times \frac{1}{2} \frac{\partial}{\partial \omega} \frac{\partial}{\partial \epsilon} \biggl\{ \biggl[ \frac{\dot{T}_r(t)}{T_r(t)} (\epsilon - \mu_r(t)) + \dot{\mu}_r(t)  \biggr]  e^{-i (\epsilon -\omega) (t_1-t)} \biggr\} .
\end{align}
Performing the integral with respect to $t_1$, and using the relations
\begin{align}
	\frac{\partial}{\partial \mu_r} f(\omega,T_r,\mu_r) &= - \frac{\partial}{\partial \omega}  f(\omega,T_r,\mu_r) , \\
	\frac{\partial}{\partial T_r} f(\omega,T_r,\mu_r) &= - \frac{\omega - \mu_r}{T_r} \frac{\partial}{\partial \omega}  f(\omega,T_r,\mu_r) .
\end{align}
on the Fermi distribution function, we obtain 
\begin{align}
&\int \! dt_1 G_s^{R,\mathrm{st.}}(t,t_1;\bm{X}(t)) \Sigma_{0,s,r}^{<,\mathrm{ad.}}(t_1,t;\bm{X}(t),\dot{\bm{X}}(t)) \nonumber \\
& = -\frac{\Gamma_r}{2} \sum_\mu \dot{X}_\mu(t) \int \frac{d\epsilon}{2\pi} \left. \frac{\partial G_s^{R,\mathrm{st.}}(\omega;\bm{X}(t))}{\partial \omega}\right|_{\omega=\epsilon} 
\nonumber \\
& \hspace{20mm} \times \frac{\partial f(\epsilon,T_r(t),\mu_r(t))}{\partial X_{\mu}}  . \label{eqn:adiabatic_current_first_term}
\end{align}
In the same manner, the integrals in the second and third terms become
\begin{align}
	&\int dt_1 dt_2 dt_3 \gamma_{s;s^{\prime}}^{R;<,\mathrm{st.}}(t,t_1;t_2,t_3;\bm{X}(t)) \nonumber \\
	&\hspace{1.5cm} \times \Sigma_{0,s}^{<,\mathrm{ad.}}(t_3,t_2;\bm{X}(t),\dot{\bm{X}}(t)) \Sigma_{0,s,r}^{<,\mathrm{st.}}(t_1,t;\bm{X}(t)) \nonumber \\
	&= i \frac{\Gamma \Gamma_r}{2}  \sum_{\mu} \dot{X}_{\mu}(t) \int \frac{d\omega_1 d\omega_2 d\epsilon}{(2\pi)^3} \nonumber \\
	& \hspace{0.5cm} \times  \left( \frac{\partial }{\partial \omega_3} - \frac{\partial }{\partial \omega_4} \right)\gamma_{s;s^{\prime}}^{R;<,\mathrm{st.}}(\omega_1,\omega_2;\omega_3,\omega_4;\bm{X}(t)) \biggr|_{\omega_3=\omega_4=\epsilon} \nonumber \\
	&\hspace{0.5cm} \times f(\omega_2,T_r(t),\mu_r(t)) \frac{\partial f_+(\epsilon,\bm{X}(t))}{\partial X_{\mu}(t)}, 
\end{align}
and
\begin{align}
&\int dt_1 dt_2 dt_3 \gamma_{s;s^{\prime}}^{<;<,\mathrm{st.}}(t,t_1;t_2,t_3;\bm{X}(t)) \nonumber \\
	&\hspace{1.5cm} \times \Sigma_{0,s}^{<,\mathrm{ad.}}(t_3,t_2;\bm{X}(t),\dot{\bm{X}}(t)) \Sigma_{0,s,r}^{A}(t_1,t) \nonumber \\
	&= i \frac{\Gamma \Gamma_r}{4}  \sum_{\mu} \dot{X}_{\mu}(t)　\int \frac{d\omega_1 d\omega_2 d\epsilon}{(2\pi)^3} \nonumber \\
	& \hspace{0.5cm} \times  \left( \frac{\partial }{\partial \omega_3} - \frac{\partial }{\partial \omega_4} \right)\gamma_{s;s^{\prime}}^{<;<,\mathrm{st.}}(\omega_1,\omega_2;\omega_3,\omega_4;\bm{X}(t)) \biggr|_{\omega_3=\omega_4=\epsilon} \nonumber \\
   &\hspace{0.5cm} \times \frac{\partial f_+(\epsilon,\bm{X}(t))}{\partial X_{\mu}(t)} ,
\end{align}
respectively.
Here $\bm{X}(t)$ is parameter vector defined in Eq.~(\ref{eq:defXvector}), and $f_{+}(\epsilon,\bm{X}(t))$ is an effective Fermi distribution function defined in Eq.~(\ref{eq:defeffectivedistribution}).
By collecting these results, the formula for the Berry connection given in Eqs.~(\ref{eqn:adiabatic_berry_connection})-(\ref{eqn:berry_connection_F_2}) can be derived.

\section{\label{app:timedelay} AC response and the Berry connection}

In this appendix, we derive Eqs.~(\ref{eqnmain:dyn_cond_result_leading}) and (\ref{eqnmain:dyn_cond_result_next_leading}) from the definition of the dynamic AC conductance, Eqs.~(\ref{eqnmain:def_dyn_cond}) and (\ref{eqnmain:def_dyn_cond2}).
We rewrite Eq.~(\ref{eqnmain:def_dyn_cond}) as
\begin{align}
	J_r(t) &= J_r^{\rm st.}({\bm X}_0) + \sum_{\mu} \int dt^{\prime} \,
    G^{\mu}_{r}(t,t^{\prime};\bm{X}_0) \delta X_{\mu} e^{-i\Omega t^{\prime}} \nonumber \\
    & \hspace{15mm} + O((\delta X_{\mu})^2), \label{eqnapp:def_dyn_cond}
\end{align}
where $G^{\mu}_{r}(t,t^{\prime};\bm{X}_0)$ is a Fourier transformation of $G^{\mu}_{r}(\Omega;\bm{X}_0)$ defined by
\begin{align}
G^{\mu}_{r}(t,t^{\prime};\bm{X}_0) = \int \frac{d\Omega}{2\pi} G^{\mu}_{r}(\Omega;\bm{X}_0) e^{-i\Omega (t-t^\prime)}.
\end{align}
To calculate the dynamic conductance $G^{\mu}_{r}(t,t^{\prime};\bm{X}_0)$ explicitly in the multi-reservoir case, we follow the procedure we have done in Sec.~\ref{sec:adiabatic_approximation}:
We pick up one of the lesser reservoir self-energies in the diagram of charge current, expand it with respect to $\delta X_{\mu}(t)$ up to the linear term, and take $\delta X_{\mu} \to 0$ limit for the rest of the reservoir self-energies.
This operation is realized by functional derivative as
\begin{align}
	&G_{r}^{\mu}(t,t^{\prime};\bm{X}_0) 
    \nonumber \\
	&= \sum_{s^{\prime},r^{\prime}} \int dt_1 dt_2 \left. \frac{\delta J_r(t)}{\delta \Sigma_{0,s^{\prime},r^{\prime}}^{<}(t_1,t_2)} \frac{\delta \Sigma_{0,s^{\prime},r^{\prime}}^{<}(t_1,t_2)}{\delta (\delta X_{\mu}(t^{\prime}))} \right|_{\delta X_{\mu} = 0} . \label{eqn:funcder_ACcond}
\end{align}
The former factor in Eq.~(\ref{eqn:funcder_ACcond}) is calculated by Eqs.~(\ref{eqn:funcder_GF_to_4point_ret}) and (\ref{eqn:funcder_GF_to_4point_les}) as
\begin{align}
	 &\left. \frac{\delta J_r(t)}{\delta \Sigma_{0,s^{\prime},r^{\prime}}^{<}(t_1,t_2)} \right|_{\delta X_{\mu} = 0} \nonumber \\
	 & \hspace{0mm}= 2e \sum_s \mathrm{Re} \Biggl[  G_s^{R,\mathrm{st.}}(t,t_1;\bm{X}_0) \delta(t_2 -t) \delta_{s,s^{\prime}} \delta_{r,r^{\prime}}  \nonumber \\
	& \hspace{15mm} + i\frac{\Gamma_r}{2} \biggl\{ \gamma_{s;s^{\prime}}^{R;<,\mathrm{st.}}(t,t;t_2,t_1;\bm{X}_0 ) \nonumber \\
    & \hspace{25mm} + \gamma_{s;s^{\prime}}^{<;<,\mathrm{st.}}(t,t;t_2,t_1;\bm{X}_0 ) \biggr\} \Biggr] .
\end{align}
The latter factor in Eq.~(\ref{eqn:funcder_ACcond}) is calculated as
\begin{align}
	&\left. \frac{\delta \Sigma_{0,s^{\prime},r^{\prime}}^{<}(t_1,t_2)}{\delta (\delta T_{r^{\prime}}(t^{\prime}))} \right|_{\delta X_{\mu} = 0} \nonumber \\
	&= i \int \frac{d\epsilon}{2\pi}\  \Gamma_{r^{\prime}} f(\epsilon,T_{r^{\prime},0},\mu_{r^{\prime},0}) e^{-i \epsilon (t_1 -t_2) } \nonumber \\
	& \hspace{5mm} \times \frac{1}{T_{r^{\prime},0}} \left ( \frac{1}{2} \delta(t_1,t,t_2)  - i (\epsilon - \mu_{r^{\prime},0})\Theta(t_1,t,t_2) \right) , \label{eqn:selfenergy_linear_AC_temp} \\
	&\left. \frac{\delta \Sigma_{0,s^{\prime},r^{\prime}}^{<}(t_1,t_2)}{\delta (\delta \mu_{r^{\prime}}(t^{\prime}))} \right|_{\delta X_{\mu} = 0} \nonumber \\
	&=  \int \frac{d\epsilon}{2\pi}\  \Gamma_{r^{\prime}} f(\epsilon,T_{r^{\prime},0},\mu_{r^{\prime},0}) e^{-i \epsilon (t_1 -t_2) }  \Theta(t_1,t,t_2) , \label{eqn:selfenergy_linear_AC_mu}
\end{align}
where
\begin{align}
	\delta(t_1,t,t_2) &= \delta(t_1-t) + \delta(t_2-t) , \\
	\Theta(t_1,t,t_2) &= \Theta(t_1-t) \Theta(t-t_2) \Theta(t_1-t_2) \nonumber \\
	& \hspace{0.5cm} -  \Theta(t_2-t) \Theta(t-t_1) \Theta(t_2-t_1) .
\end{align}
To calculate low frequency limit of the dynamic conductance $G_r^{\mu}(\Omega;\bm{X}_0)$, let us consider Fourier transformation of Eqs.~(\ref{eqn:selfenergy_linear_AC_temp}) and (\ref{eqn:selfenergy_linear_AC_mu}), and expand them with respect to frequency.
From Eq.~(\ref{eqn:selfenergy_linear_AC_temp}), we obtain 
\begin{align}
	&\int dt^{\prime} \ e^{-i\Omega t^{\prime}} \left. \frac{\delta \Sigma_{0,s^{\prime},r^{\prime}}^{<}(t_1,t_2)}{\delta (\delta T_{r^{\prime}}(t^{\prime}))} \right|_{\delta X_{\mu} = 0}   \nonumber \\
	&= \delta_{T_{r^{\prime}}} \Sigma_{0,s^{\prime},r^{\prime}}^{<,(0)}(t_1,t_2) - i \Omega \delta_{T_{r^{\prime}}} \Sigma_{0,s^{\prime},r^{\prime}}^{<,(1)}(t_1,t_2) + O(\Omega^2) , \label{eqn:selfenergy_linear_AC_temp_fourier} \\
	& \delta_{T_{r^{\prime}}} \Sigma_{0,s^{\prime},r^{\prime}}^{<,(0)}(t_1,t_2) = \int dt^{\prime} \  \left. \frac{\delta \Sigma_{0,s^{\prime},r^{\prime}}^{<}(t_1,t_2)}{\delta (\delta T_{r^{\prime}}(t^{\prime}))} \right|_{\delta X_{\mu} = 0} \nonumber \\
	& = \frac{\partial \Sigma_{0,s^{\prime},r^{\prime}}^{<,\mathrm{st.}}(t_1,t_2;\bm{X}_0)}{\partial T_{r^{\prime}}} , \\
	&i \Omega \delta_{T_{r^{\prime}}} \Sigma_{0,s^{\prime},r^{\prime}}^{<,(1)}(t_1,t_2) = \int dt^{\prime} \ i\Omega t^{\prime} \left. \frac{\delta \Sigma_{0,s^{\prime},r^{\prime}}^{<}(t_1,t_2)}{\delta (\delta T_{r^{\prime}}(t^{\prime}))} \right|_{\delta X_{\mu} = 0} \nonumber \\
	&=  - \Omega \int \frac{d\epsilon}{2\pi}\  \Gamma_{r^{\prime}} f(\epsilon,T_{r^{\prime},0},\mu_{r^{\prime},0}) e^{-i \epsilon (t_1 -t_2) } \frac{1}{2} (t_1 + t_2)\nonumber \\
	& \hspace{1.5cm} \times \frac{1}{T_{r^{\prime},0}} \left[ 1 - i (\epsilon - \mu_{r,0})  (t_1- t_2) \right] . \label{eqn:dyn_cond_temp_next}
\end{align}
From Eq.~(\ref{eqn:selfenergy_linear_AC_mu}), we obtain 
\begin{align}
	&\int dt^{\prime} \ e^{-i\Omega t^{\prime}} \left. \frac{\delta \Sigma_{0,s^{\prime},r^{\prime}}^{<}(t_1,t_2)}{\delta (\delta \mu_{r}(t^{\prime}))} \right|_{\delta X_{\mu} = 0} \nonumber \\
	&=  \delta_{\mu_{r^{\prime}}} \Sigma_{0,s^{\prime},r^{\prime}}^{<,(0)}(t_1,t_2) - i \Omega \delta_{\mu_{r^{\prime}}} \Sigma_{0,s^{\prime},r^{\prime}}^{<,(1)}(t_1,t_2) + O(\Omega^2),  \label{eqn:selfenergy_linear_AC_mu_fourier} \\
	&\delta_{\mu_{r^{\prime}}} \Sigma_{0,s^{\prime},r^{\prime}}^{<,(0)}(t_1,t_2) = \int dt^{\prime} \  \left. \frac{\delta \Sigma_{0,s^{\prime},r^{\prime}}^{<}(t_1,t_2)}{\delta (\delta \mu_{r^{\prime}}(t^{\prime}))} \right|_{\delta X_{\mu} = 0} \nonumber \\
	&= \frac{\partial \Sigma_{0,s^{\prime},r^{\prime}}^{<,\mathrm{st.}}(t_1,t_2;\bm{X}_0)}{\partial \mu_{r^{\prime}}} , \\
	&i \Omega \delta_{\mu_{r^{\prime}}} \Sigma_{0,s^{\prime},r^{\prime}}^{<,(1)}(t_1,t_2) =\int dt^{\prime} \ i\Omega t^{\prime} \left. \frac{\delta \Sigma_{0,s^{\prime},r^{\prime}}^{<}(t_1,t_2)}{\delta (\delta \mu_{r}(t^{\prime}))} \right|_{\delta X_{\mu} = 0} \nonumber \\
	&= i \Omega \int \frac{d\epsilon}{2\pi}\  \Gamma_r f(\epsilon,T_{r,0},\mu_{r,0}) e^{-i \epsilon (t_1 -t_2) } (t_1^2 - t_2^2)  .  \label{eqn:dyn_cond_mu_next}
\end{align}
Substituting Eqs.~(\ref{eqn:selfenergy_linear_AC_temp_fourier})-(\ref{eqn:dyn_cond_mu_next}) into Eq. (\ref{eqn:funcder_ACcond}), we can expand $G_r^{\mu}(\Omega;\bm{X}_0)$ with respect to $\Omega$ as
\begin{align}
	G_r^{\mu}(\Omega;\bm{X}_0) = G_{r,0}^{\mu}(\bm{X}_0) - i \Omega G_{r,1}^{\mu}(\bm{X}_0) +  O(\Omega^2) ,
\end{align}
where, for $n \in \{0,1\}$, 
\begin{align}
	&G_{r,n}^{\mu}(\bm{X}_0) \nonumber \\
	&= \sum_{s^{\prime},r^{\prime}} \int dt_1 dt_2 \ \left. \frac{\delta J_r(t)}{\delta \Sigma_{0,s^{\prime},r^{\prime}}^{<}(t_1,t_2)} 
    \right|_{\delta X_{\mu} = 0} \!\!\! \delta_{X_{\mu}} \Sigma_{0,s^{\prime},r^{\prime}}^{<,(n)}(t_1,t_2) .
\end{align}
Comparing Eqs.~(\ref{eqn:dyn_cond_temp_next}) and (\ref{eqn:dyn_cond_mu_next}) with Eq.~(\ref{eqn:adiabatic_reseroir_lesser_selfenergy}), we finally conclude that coefficients $G_r^{\mu, n}(\bm{X}_0)$ are related by Eqs.~(\ref{eqnmain:dyn_cond_result_leading}) and (\ref{eqnmain:dyn_cond_result_next_leading}).

\section{\label{app:RenormalizedParameters}Renormalized parameters}

The renormalized parameters are calculated by following relations~\cite{Hewson01}:
\begin{align}
	n_{d} &= \frac{1}{2} - \frac{1}{\pi} \arctan (2 \tilde{\epsilon}_d / \tilde{\Gamma}) ,\label{eqn:phaseshift} \\
	\chi_s &= \frac{(g \mu_b)^2}{2} \tilde{\rho} (1 + \tilde{U} \tilde{\rho} ) , \\
	\chi_c &= 2 \tilde{\rho} (1 - \tilde{U} \tilde{\rho} ) , \label{eqn:chi_c}
\end{align}
where $n_d$, $\chi_s$ and $\chi_c$ denote the occupation, spin susceptibility and charge susceptibility of the electron in the QD, respectively.
$g$ is the g-factor and $\mu_b$ is the Bohr magneton.
$\tilde{\rho}$ is renormalized density of states of electron in the QD at the Fermi level, defined as
\begin{align}
	\tilde{\rho} = \frac{\tilde{\Gamma}/2\pi}{\tilde{\epsilon}_d^2 + \tilde{\Gamma}^2/4} .
\end{align}
To compute renormalized parameters by Eqs.~(\ref{eqn:phaseshift})-(\ref{eqn:chi_c}), the three quantities, $n_d$, $\chi_s$ and $\chi_c$, are calculated by Bethe Ansatz~\cite{Okiji82,Wiegmann83}.

\bibliography{references}

\end{document}